%% file: power-techreport.tex
\documentclass[conference]{IEEEtran}
\usepackage{cite}
\usepackage{epsfig}
\usepackage{color}
\usepackage{amsmath}
\usepackage{amssymb}
\usepackage{amsfonts}
\usepackage{graphicx}
\usepackage{subfigure}
\usepackage{algorithm}
\usepackage{url}

\usepackage{algpseudocode}
\usepackage{fancyhdr}

\usepackage{multirow}

\usepackage{etoolbox}

\catcode`~=11 
\newcommand{\urltilde}{\kern -.15em\lower .7ex\hbox{~}\kern .04em}
\catcode`~=13 

\algdef{SE}[FUNCTION]{RFunc}{EndRFunc}%
   [3]{#1 \algorithmicfunction\ \textproc{#2}\ifthenelse{\equal{#3}{}}{}{(#3)}}%
   {\algorithmicend\ \algorithmicfunction}%
\algdef{SE}[FUNCTION]{RProc}{EndRProc}%
   [4]{#1 \algorithmicprocedure\ \textproc{#2}\ifthenelse{\equal{#3}{}}{}{(#3)\ #4}}%
   {\algorithmicend\ \algorithmicprocedure }%

\input{./my_includes.tex}

\newtheorem{theorem}{Theorem}[section]

\newtheorem{corollary}[theorem]{Corollary}
\newtheorem{lemma}[theorem]{Lemma}

\newtheorem{definition}{Definition}

\newtheorem{observation}{Observation}[section]

\def\eod{\vrule height 6pt width 5pt depth 0pt}

\newcommand{\reals}{\mathbb R}

\newcommand{\remove}[1]{}

\IEEEoverridecommandlockouts

\begin{document}



\pagenumbering{arabic} \setcounter{page}{1}

\title{Power Grid Vulnerability to Geographically Correlated
  Failures -- \\ Analysis and Control Implications}

\author{Andrey Bernstein\IEEEauthorrefmark{1}\IEEEauthorrefmark{2},
        Daniel Bienstock\IEEEauthorrefmark{3},
        David~Hay\IEEEauthorrefmark{4}, 
        Meric Uzunoglu\IEEEauthorrefmark{1},
        and~Gil~Zussman\IEEEauthorrefmark{1}\\
\IEEEauthorrefmark{1}Department of Electrical Engineering, Columbia University, New York, NY 10027 \\
\IEEEauthorrefmark{2}Department of Electrical Engineering, Technion, Haifa 32000, Israel, \\
\IEEEauthorrefmark{3}Department of Applied Physics and Applied Math,  Columbia University, New York, NY 10027 \\
\IEEEauthorrefmark{4}School of Engineering and Computer Science, Hebrew University, Jerusalem 91904, Israel \\
Email:  andreyb@techunix.technion.ac.il,  dano@columbia.edu, dhay@cs.huji.ac.il, \\ meric.uzunoglu@gmail.com, gil@ee.columbia.edu
}

%

\maketitle

\thispagestyle{fancy} 

\input{./abstract_intro.tex}

\input{./related.tex}

\input{./model.tex}
\input{./theory.tex}

\input{./algos.tex}

\input{./data.tex}
\input{./results.tex}

\input{./sandiego.tex}

\input{./control.tex}

\input{./conc.tex}

\section*{Acknowledgements}
This work was supported in part by DOE award DE-SC000267, the Legacy
Heritage Fund program of the Israel Science Foundation (Grant No.\ 1816/10),
DTRA grant HDTRA1-09-1-0057, a grant from the from the U.S.-Israel Binational Science Foundation, and NSF grants CNS-10-18379 and CNS-10-54856.
We thank Eric Glass for help with GIS data processing.

\bibliographystyle{IEEEtranS}
\bibliography{power}
\input{./appendix.tex}
\end{document}

%% file: my_includes.tex
\renewcommand{\l}{\left}
\renewcommand{\r}{\right}

\DeclareSymbolFont{AMSb}{U}{msb}{m}{n}
\DeclareMathSymbol{\N}{\mathbin}{AMSb}{"4E}
\DeclareMathSymbol{\ZZ}{\mathbin}{AMSb}{"5A}
\DeclareMathSymbol{\Y}{\mathbin}{AMSb}{"59}
\DeclareMathSymbol{\U}{\mathbin}{AMSb}{"55}
\DeclareMathSymbol{\RR}{\mathbin}{AMSb}{"52}
\DeclareMathSymbol{\Q}{\mathbin}{AMSb}{"51}
\DeclareMathSymbol{\Prob}{\mathbin}{AMSb}{"50}
\DeclareMathSymbol{\HHH}{\mathbin}{AMSb}{"48}
\DeclareMathSymbol{\I}{\mathbin}{AMSb}{"49}
\DeclareMathSymbol{\C}{\mathbin}{AMSb}{"43}
\DeclareMathSymbol{\E}{\mathbin}{AMSb}{"45}
\DeclareMathSymbol{\C}{\mathbin}{AMSb}{"43}
\DeclareMathSymbol{\D}{\mathbin}{AMSb}{"44}
\DeclareMathSymbol{\OO}{\mathbin}{AMSb}{"4F}
\DeclareMathSymbol{\X}{\mathbin}{AMSb}{"58}
\DeclareMathSymbol{\LL}{\mathbin}{AMSb}{"4C}

\newcommand{\Nl}{\mathcal{N}}

\newcommand{\Cl}{\mathcal{C}}

\newcommand{\Gl}{\mathcal{G}}
\newcommand{\Dl}{\mathcal{D}}

\newenvironment{Rems}[1][Remarks.]{\begin{trivlist}
\item[\hskip \labelsep {\bfseries #1}]}{\end{trivlist}}

%% file: abstract_intro.tex
\begin{abstract}
We consider power line outages in the \emph{transmission system} of the power grid, and specifically those caused by a natural disaster or a large scale physical attack. In the transmission system, an outage of a line may lead to overload on other lines, thereby eventually leading to their outage. While such cascading failures have been studied before, our focus is on cascading failures that follow an \textit{outage of several lines in the same geographical area}. We provide an analytical model of such failures, investigate the model's properties, and show that it differs from other models used to analyze cascades in the power grid (e.g., epidemic/percolation-based models). We then show how to identify the most vulnerable locations in the grid and perform extensive numerical experiments with real grid data to investigate the various effects of geographically correlated outages and the resulting cascades. These results allow us to gain insights into the relationships between various parameters and performance metrics, such as the size of the original event, the final number of connected components, and the fraction of demand (load) satisfied after the cascade. In particular, we focus on the timing and nature of {\em optimal control} actions
used to reduce the impact of a cascade, in real time.  We also compare results obtained by our model to the results of a real cascade that occurred during a major blackout in the San Diego area on Sept. 2011. The analysis and results presented in this paper will have implications both on the design of new power grids and on identifying the locations for shielding, strengthening, and monitoring efforts in grid upgrades.

\end{abstract}

\begin{IEEEkeywords}
Power Grid, Geographically-Correlated Failures, Cascading Failures, Resilience, Survivability.
\end{IEEEkeywords}





\section{Introduction}
\label{sec:intro}

Recent colossal failures of the power grid (such as the Aug. 2003 blackout in the
Northeastern United States and Canada~\cite{Blackout, SA_Amin}) demonstrated that large-scale and/or long-term failures will
have devastating effects on almost every aspect in modern life, as well as on
interdependent systems (e.g., telecommunications, gas and water supply, and transportation). Therefore, there is a need  the study the vulnerability
of the existing power transmission system and to identify ways to mitigate large-scale blackouts.

The power grid is vulnerable to natural disasters, such as earthquakes, hurricanes, floods, and solar flares \cite{EMP_flare} as well as to physical attacks, such as an Electromagnetic Pulse (EMP) attack \cite{EMP,EMP_flare, OneSecond}. Thus, we focus the vulnerability of the power grid to an outage of several lines in the same geographical area (i.e., to \emph{geographically-correlated failures}). 
Recent works focused on identifying a \emph{few
vulnerable lines} throughout the entire network \cite{Dan, Dan2,pinar_power}, on designing line or node interdiction strategies \cite{BGHMW07,SWB09},  and  on characterizing the network graph and studying probabilistic failure propagation models \cite{vermont,yeh,yeh2,Dobson,Chassin2005667,Wang}. 
However, our objective is to identify \emph{the most vulnerable areas}
in the power grid, and examine appropriate real-time control countermeasures to minimize the impact of an event of this type. Detection of the most
vulnerable areas has various practical applications, since the system in these areas can be either
shielded (e.g., against EMP attacks or solar flares), strengthened (e.g., by increasing the capacities of some relevant lines), or
monitored (e.g., as part of smart grid upgrade projects). Real-time control
will likely still be needed, in case the pre-strengthening of the grid has
overlooked a particular ``attack'' pattern.


\begin{figure}
\centering
\epsfig{figure=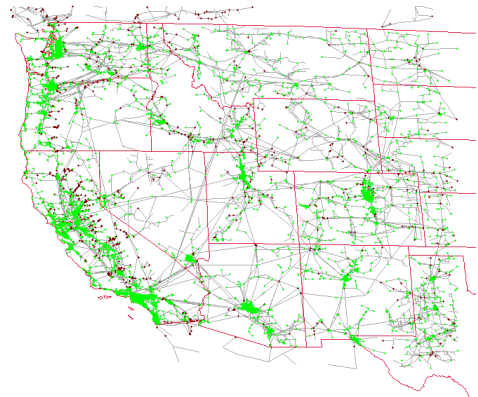, width=0.35\textwidth}
\vspace*{-1em}
\caption{\label{fig:network}The graph that represents part of the Western Interconnect and sections of the Texas, Oklahoma, Kansas, Nebraska, and the Dakotas' grids. Green dots are demands (loads), red dots are supplies (generators), and neutral nodes are not shown. This graph was used to derive numerical results.}
\end{figure}

Unlike graph-theoretical network flows, the flow in the power grid is governed by the laws of physics and there are \emph{no strict capacity bounds on the lines}~\cite{Model, Dobson,low_power}. On the other hand, there is a \emph{rating threshold} associated with each line, such that when the flow through a line exceeds the threshold, the line heats up and eventually faults. Such an outage, in turn, causes another change in the power grid, that can eventually lead to a \emph{cascading failure}. We describe the \emph{linearized (i.e., DC) power flow model} and the \emph{cascading failure model} (originated from \cite{Chen2005318} and extended in \cite{Dan,Dan2}) that allow us to obtain analytic and numerical results despite the problem's complexity.  We note that severe cascading failures are \emph{hard to control in real time} \cite{Dan, Dan2, Dobson,AWM07,PTC11}, since the power grid optimization and control problems are of enormous size. Thus, in our numerical experiments, we apply basic control mechanisms that shed demands at the end of each round.



We initially use the linearized power flow model to study some simple motivating examples and to derive analytical results regarding the cascade propagation in simple ring-based topologies. We note that several previous works (e.g., \cite{yeh,yeh2,Chassin2005667,SmartGridComplex, Havlin11} and references therein) assumed that a line or node failure leads with some probability  to a failure of nearby nodes or lines. Such epidemic-based modeling allows using percolation-based tools to analyze the effects of a cascade. However,  we show that using the more realistic power flow model leads to failure propagation characteristics that are significantly different. Specifically, we show that a failure of a line can lead to a failure of a line $M$ hops away (with $M$ arbitrarily large). This result is of particular importance, since it has been observed in real cascades (as the recent one in the San Diego area, discussed in Section \ref{sec:SanDiego}). Moreover, we prove that cascading failures can last arbitrarily long time and that a network whose topology is a subgraph of another topology can be more resilient to failures.
\pagestyle{plain}

In this paper, we focus on contingency events in a grid that are initiated by geographically correlated failures. We represent the area affected by a
contingency by a disk. Since such a disk  can  theoretically be placed in an infinite number of locations, we briefly discuss an efficient computational geometric method (which builds on results from \cite{infocom11}) that allows identifying a finite set of locations that includes all possible failure events.

In our numerical experiments, we use the WECC (Western-Interconnect) real power grid data taken from  the Platts Geographic Information System (GIS) \cite{GIS} (the resulting graph appears in Figure~\ref{fig:network}).
We present extensive numerical results obtained by simulating the cascading failures for each of the possible disk centers (the results have been obtained by repeatedly and efficiently solving very large systems of equations). When only few lines initially fault, cascading failures usually start slowly and intensify over time \cite{Dan,Dan2}. However, when the failures are geographically correlated, we notice that in many cases this \emph{slow start} phenomenon does not exist. 
We illustrate the effects of the most (and the least) devastating failures and show the yield (the overall reduction in power generation) for all different failure locations in the Western US. Moreover, we identify various relations between parameters and performance metrics (such as yield, number of components into which the network partitions, and number of faulted lines which corresponds to the length of the repair process). We also study the sensitivity of the results to different failure models (stochastic vs.\ deterministic model), to the value of the Factor of Safety (the ratio between line capacity and normal flow on the line), and to whether or not the line capacities are derived based on $N-1$ contingency plan.

While large scale cascades are quite rare \cite{Chen2005318}, during our work on this report, a cascade took place in the San Diego area (on Sept. 8, 2011). The causes of this cascade are still under investigation. Yet, we have been able to use some preliminary information \cite{SD_briefing} to assess the accuracy of our method and parameters. We conclude by discussing some of the numerical results that we obtained for a similar scenario and demonstrating that our methods have the potential to identify vulnerable parts of the network.

Finally, we consider optimal control actions to be taken in the event of
such a failure. 
We show, experimentally, that appropriate action
taken at the appropriate time (and not necessarily at the start of the cascade)
can rapidly stop the process, while losing a minimum quantity of demand.

Below are the main contributions of this paper. First, we obtain analytical results regarding network vulnerability and resilience under the power flow model which is significantly different from the classical network flow models. These results provide \emph{insights that significantly differ from insights obtained from epidemic/percolation-based models}. Second, we combine techniques from optimization, computational geometry, and communications network vulnerability analysis to develop a method that allows obtaining extensive numerical results regarding the effects of geographically correlated failures on a real grid. To the best of our knowledge, \emph{this is the first attempt to obtain such results using real geographical data}.  Third, we briefly illustrate that control actions have the potential to mitigate the effects of even the worst case failure. The results obtained in this paper will provide insight into the design of control algorithms and network architectures.

The rest of the paper is organized as follows. In Section~\ref{sec:related} we present the related work and in Section~\ref{sec:power_mod} we describe the power flow and cascade models. Section~\ref{sec:theory} provides analytical results regarding simple network topologies. Section~\ref{sec:res} presents the method used to set some of the parameters and the algorithm used to identify the most vulnerable locations within the grid. Section~\ref{sec:data} describes the power grid data used in this paper . In Section~\ref{sec:results} we present our numerical findings regarding large scale failure and in Section \ref{sec:SanDiego} we discuss the recent cascade in the San Diego area. Section ~\ref{sec:control} describes optimal control methods and their impact in the particular case of one of
the simulated worst-case events  that we compute. Section~\ref{sec:conc} provides concluding remarks and directions for future work.

%% file: related.tex
\section{Related Work} \label{sec:related}

The power grid and its robustness have drawn a lot of attention recently,
as efforts are being made to create a smarter, more efficient, and
more sustainable power grid (see, e.g.,~\cite{SA_Amin,ProcIEEE}). The power grid is traditionally modeled as a complex system,
made up of many components, whose interactions (e.g., power
flows and control mechanisms) are not effectively
computable (e.g.,~\cite{Model,GoranAderson} and
references therein).



When investigating the robustness of the power grid, cascading failures are a
major concern~\cite{HBC09, Dobson, SA_Amin,Chen2005318}.  This phenomenon, along with
other vulnerabilities of the power grid,  was studied from a few different viewpoints. First, several papers studied common topological properties of  power grid networks and probabilistic failure propagation models, so that one can evaluate the behavior of a generic grid as a self-organized critical
system using, for example, \emph{percolation theory} (see \cite{Motter1,SmartGridComplex, yeh, yeh2,Chassin2005667,vermont,Wang,D15,Havlin11} and references therein). These works are closely related to a long line of research in the power community which uses  monte carlo techniques to analyze system reliability (e.g., \cite{BL94})
  

Another major line of research focused on specific (microscopic) power flow models and used them to identify a \emph{few
vulnerable lines} throughout the entire network \cite{Dan, Dan2,pinar_power}. In particular, \cite{BGHMW07,SWB09} focus on designing line or node interdiction strategies that will lead to an effective attack on the grid. Since the problem is computationally intractable,  most of these
papers use a linearized direct-current (DC) model, which is a
tractable relaxation of the exact alternating-current (AC) power flow model.
In addition, the initial failure events (causing eventually the cascading failures) are assumed to be sporadic link
outages (and in most cases, a single link outage), with no correlation between them.
On the contrary, we focus on events that cause \emph{a large number of
failures in a specific geographical region}
(e.g.,~\cite{EMP,EMP_flare}). To the best of our knowledge, geographically-correlated
failures in the power grid have not been considered before and have been studied only recently in the context of communication
networks~\cite{G28,Seb,infocom11, my_thai,S09}. 

Moreover, since cascading failures are highly dependent
on the network topology, we use the real topology of the
western U.S. power grid. While building on the linearized model, we manage to obtain numerical results for a very large scale real network (results for large networks have been derived in the past using mostly probabilistic models \cite{Chassin2005667,vermont,SmartGridComplex}). Perhaps, closest to us in its approach is \cite{PTC11} that analyzes the effects of single line failures on the Polish power grid. Yet, while \cite{PTC11}  considers a real topology, it does not take into account geographical effects. Moreover, it  applies control mechanisms that are more sophisticated than the ones considered here. Control mechanisms are also introduced in \cite{AWM07} that develops a method to trade off load shedding  and cascade propagation risk.
Recently, \cite{low_power, low2} proposed efficient optimization algorithms for solving the classical power flow problems. These algorithms use efficient mathematical programming methods and can support \emph{offline} control decisions. However, they are not applicable when rapid \emph{online} control of the grid is required.

%% file: model.tex
\section{Basic Models}
\label{sec:power_mod}


We adopt the \emph{linearized} (or DC) \emph{power flow model}, which is widely used as an approximation for more realistic non-linear AC power model (see~\cite{Model} for a survey on the power flow models).
In particular, we follow~\cite{Dan,Dan2} and represent the power grid by a directed graph $\Gl$, whose set of nodes is $\Nl$. Each of these nodes is classified either as a \emph{supply node} (``generator''), a \emph{demand node} (``load''), or a \emph{neutral node}. Let $\Dl \subseteq \Nl$ be the set of the demand nodes, and for each node $i \in \Dl$, let $D_i$ be its demand. Also, $ \Cl \subseteq \Nl$ denotes the set of the supply nodes and for each node $i \in \Cl$, $P_i$ is the active power generated at $i$. The edges of the graph $\Gl$ represent the transmission lines. The orientation of the lines is arbitrarily and is simply used for notational convenience.  We also assume \emph{pure reactive} lines, implying that each line $(i, j)$ is characterized by its \emph{reactance} $x_{ij}$.


Given supply and demand vectors $(P, D)$, a \emph{power flow} is a solution $(f, \theta)$ of the following system of equations:
\begin{eqnarray}
\label{eqn:flow1}&&\sum_{(i, j) \in \delta^+(i)} f_{ij} - \sum_{(j, i) \in \delta^-(i)} f_{ji} =
\begin{cases}
P_i, & i \in \Cl \\
-D_i, & i \in \Dl \\
0, & \text{otherwise}
\end{cases} \\
\label{eqn:flow2}&&\theta_i - \theta_j - x_{ij}f_{ij} = 0, \ \ \forall (i, j)
\end{eqnarray}
where $\delta^+(i)$ ($\delta^-(i)$) is the set of lines oriented out of (into) node $i$,
$f_{ij}$ is the (real) power flow along line $(i,j)$, and $\theta_{i}$ is the phase angle of node $i$.
These equations guarantee power flow conservation 
in each neutral node, and take into account the reactance of each line. In addition, since the orientation of lines is arbitrary, a negative flow value simply means a flow in the opposite direction.

When $\Gl$ is fully connected and $\sum_{i \in \Cl} P_i = \sum_{i \in \Dl} D_i$, (\ref{eqn:flow1})--(\ref{eqn:flow2}) has a unique solution~\cite[Lemma 1.1]{Dan}. This holds even when $\Gl$ is not connected but the total supply and demand within each of the connected components are equal.

We note that the DC power flow model resembles an \emph{electrical circuit model}, where phase angles are analogous to voltages, reactance is analogous to resistance, and the power flow is analogous to the current.
The following observation, which is analogous to Kirchoff's law, captures the essence of the model and provides easier way to look at (\ref{eqn:flow1})--(\ref{eqn:flow2}) analytically. It uses the notion of a path between nodes,
which is an alternating sequence of lines and adjacents nodes. (Proof is in the appendix).
\begin{observation}
\label{lem:samesum}
Consider two nodes $a$ and $b$ and two paths
$\pi_1=(a,e_0,v_0,e_1,v_1,\ldots, b)$ and
$\pi_2=(a,e_0',v_0',e_1',v_1',$ $\ldots, b)$. The sum of the flow-reactance product $f_{e_i}x_{e_i}$  along
the lines of path $\pi_1$ is equal to the sum of the flow-reactance product  along
the lines of path $\pi_2$. Specifically, the flows
along parallel lines with the same reactance is the same.
\end{observation}

\begin{algorithm}[t]
\textbf{Cascading Failure Model (Deterministic Case)}\vspace{1mm}
{\hrule height0.8pt depth0pt \kern0pt}\vspace{-.5mm}
\small
\begin{trivlist}
\item \textbf{Input:} Connected network graph $\Gl$.
\item \textbf{Initialization:} Before time step $t = 0$, we have that $\sum_{i \in \Cl} P_i = \sum_{i \in \Dl} D_i$ (i.e.,  the power is balanced), (\ref{eqn:flow1})--(\ref{eqn:flow2}) are satisfied for $\Gl$, and all flows along all lines are within the corresponding power capacity.
\item \textbf{Failure event:}  At time step $t = 0$, a failure of some subset of links of $\Gl$ occurs. Let $\Gl\mbox{.changed} = \mbox{true}$.
\item \textbf{While $\Gl\mbox{.changed is true}$ do:}
\begin{enumerate}
\item Adjust the total demand to the total supply \emph{within each component} of $\Gl$.
\item Use the system  (\ref{eqn:flow1})--(\ref{eqn:flow2}) to recalculate the power flow in $\Gl$.
\item For all lines compute a moving average $$\tilde{f}^t_{ij}=\alpha|f_{ij}|+ (1-\alpha)\tilde{f}^{t-1}_{ij}$$
\item Remove from $\Gl$ all lines with  flow \emph{moving average} above power capacity ($\tilde{f}^t_{ij} > u_{ij}$). If at least one line was removed at this step, let $\Gl\mbox{.changed} = \mbox{true}$; otherwise,  let $\Gl\mbox{.changed} = \mbox{false}$.
\end{enumerate}
\end{trivlist}
\normalsize
\end{algorithm}

Next we describe the \emph{Cascading Failure Model}.
We assume that each line $(i,j)$  has a predetermined \emph{power capacity}  $u_{ij}$, which bounds its power flow in a normal operation of the system (that is, $|f_{ij}|\leq u_{ij}$).
We assume that before a failure event, $\Gl$ is fully connected, the total supply and demand are equal, the power flows satisfy (\ref{eqn:flow1})--(\ref{eqn:flow2}), and the power flow of each line is at most its power capacity. Upon a failure, some lines are removed from the graph, implying that it may become disconnected. Thus, within each component, we adjust the total demand to equal  the total supply, by decreasing the demand (supply) by the same factor at all loads (generators). This process is sometimes  called \emph{demand shedding} and naturally it causes a decrease in demand/supply. Then, we
use (\ref{eqn:flow1})--(\ref{eqn:flow2}) to recalculate the power flows in the new graph. The new flows \emph{may exceed the capacity} and as a result, the corresponding lines will become overheated. Thermal effects cause overloaded lines to become more sensitive to a large number of effects each of which could cause failure.  We model outages using a \emph{moving average} of the power flow, using a value $\tilde{f}^t_{ij}=\alpha |f_{ij}|+ (1-\alpha)\tilde{f}^{t-1}_{ij}$ (in this paper, we mostly use either $\alpha=0.5$ or $\alpha=1$).  To first order, this approximates thermal effects, including heating and cooling from prior states.  A similar moving average model was considered in~\cite{AWM07, Dan}.
%
 A general \emph{outage rule} gives the fault probability of line $(i,j)$, given its moving average $\tilde{f}^t_{ij}$. In this paper, we consider the following rule:
\begin{equation} \label{eqn:outage}
\Prob \l\{ \text{Line } (i, j) \text{ faults at round } t \r\} =
\begin{cases}
1, & \tilde{f}^t_{ij} > (1 + \varepsilon) u_{ij} \\
0, & \tilde{f}^t_{ij} \leq (1 - \varepsilon) u_{ij} \\
p, & \text{ otherwise}.
\end{cases}
\end{equation}
where $0 \leq \varepsilon < 1$  and $0 \leq p \leq 1$  are parameters. When $\varepsilon = 0$, we obtain a \emph{deterministic} version of this rule. In this case, lines $(i,j)$  whose $\tilde{f}^t_{ij}$ is above the power capacity $u_{ij}$ are removed from the graph.

The process is repeated 
in rounds \emph{until the system reaches stability}, namely until there is an iteration in which no lines are removed. 
%
We note that our model does not have a notion of exact time, however the relation between the elapsed time and the corresponding time can be adjusted by using different values of $\alpha$; in a sense, up to a certain degree, smaller value of $\alpha$ implies that we take a more microscopic look at the cascade.

Our major metric to assess the severity of a cascading failure is the system post-failure \emph{yield} which is defined as follows:
\begin{equation} \label{eqn:tp}
Y \triangleq \frac{\text{The actual demand at the stability}}{\text{The original demand}}.
\end{equation}

In addition to the yield, other performance metrics will be considered, such as the number of faulted lines, the number of connected components, and the maximum line overload. While the yield naturally gives an assessment of the severity of the cascade after the process has already finished, the other metrics may also shed light on the cascade properties after a fixed (given) number of rounds.

Note that our model contains a very simple control mechanism, namely, round-by-round demand shedding. In Section~\ref{sec:control}, we consider more elaborated control mechanisms and compare the results.

%% file: theory.tex
\section{Cascading Failures Properties in Simple Graphs}
\label{sec:theory}
In this section we describe important properties of the power flow and
cascading failure
models in a simple graph.
Specifically, we show that unlike other flow models,
\emph{cascading failures in a power flow
network are harder to predict and are different from epidemic-like
failure models}, from the following four reasons:
\begin{enumerate}
\item
Consecutive failures in a cascade may happen within an arbitrarily
long distance of each other.
\item
Cascading failures can last arbitrarily long time.
\item
A network $\mathcal{G}_1$ whose topology is a subgraph of another network $\mathcal{G}_2$ can be
more resilient to failures than $\mathcal{G}_2$.
\item
A failure event which results in initial
failure of some set of lines $A$ can cause more damage than a failure
event, whose initial set of faulted lines is a superset of $A$.
\end{enumerate}

\begin{figure}
\centering
\epsfig{figure=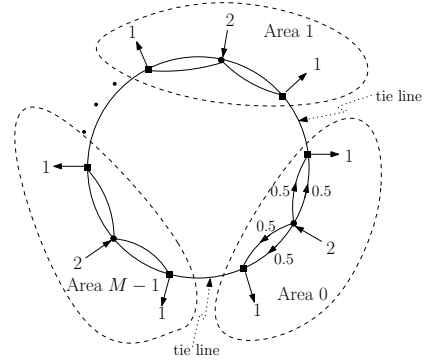, width=0.3\textwidth}
\caption{The $M$-ring $\mathcal{R}_M$. Each generation node and its
  two adjacent demand node are a self-sustained area. No power flow
  is transmitted between these areas (that is, along the tie lines). \label{fig:areas}}
\end{figure}

In order to prove the results, we use the following simple power flow
topology, which is depicted in Figure~\ref{fig:areas}

\begin{definition}
An $M$-ring $\mathcal{R}_M=\langle \Cl\cup\Dl,
\mathcal{E} \rangle$ is a directed graph with $M$ supply
nodes $\Cl=\{0,\ldots,M-1\}$, $2M$ demand nodes
$\Dl=\{M,\ldots,3M-1\}$,  two parallel transmission lines connecting each
generator  $i\in\Cl$ to demand node $M+2i$,
two parallel transmission lines connecting generator $i\in\Cl$ to demand
node $M+2i+1$, and a single transmission line connecting demand node
$M+2i+1$ ($i\in\Cl$)  to demand node $M+(2i+2 \mod
2M)$\footnote{Parallel transmission lines are denoted $(i,j)$ and
  $(i,j)'$. The orientation of the lines is only for notational
  convenience.}. For each $i\in{\Cl}$, the generation value is $P_i=2$,
and the demand values are $D_{2i}=D_{2i+1}=1$. The reactance value $x_{ij}$ of all
transmission lines is $1$.
\end{definition}

Clearly, one can view the $M$-ring as a collection of $M$ self
sustained areas: each generator $i$ supplies the demands of
$M+2i$ and $M+2i+1$.  For brevity, we call the lines connecting
$i$ and $M+2i$ \emph{even lines}, the lines connecting $M+2i+1$
\emph{odd lines}, and the lines connecting two demands (that is,
connecting two self-sustained areas) \emph{tie lines}. Moreover, we
refer to odd and even lines collectively as \emph{internal
  lines}.

By symmetry, we have the same power flow on all even lines and odd
lines.
The solution of (\ref{eqn:flow1})--(\ref{eqn:flow2}) verifies that this is indeed the
case: a power flow of $0.5$ is transmitted from each generator along
its $4$ adjacent internal lines, as shown in Figure~\ref{fig:areas}, the phase angle of all
generators is the same, and the phase angle of all demand
nodes is the same. This also implies that there is no flow on
the tie lines. Notice that if all lines has a power capacity of
$0.5$, it suffices to sustain that power flow.

In the rest of the section, we will consider the following types
of failure events:
\begin{itemize}
\item \textbf{Area failure:} The four internal lines within
  Area $i$, as well as the two lines connecting Area $i$ to the
  adjacent areas, fault. Namely, lines
  $(i,M+2i),(i,M+2i)',(i,M+2i+1),(i,M+2i+1)', (M+(2i-1\mod
  2M),M+2i),(M+2i+1,M+(2i+2\mod 2M))$ fault.
\item \textbf{Parallel lines failure:} Two parallel internal lines within
  Area $i$ fault. Without loss of generality, we consider only
  even parallel lines,  that is, $(i,M+2i)$ and $(i,M+2i)'$ fault.
\item \textbf{Odd line and even line failure:} One odd  line and one even line within
  Area $i$ fault. Namely, lines $(i,M+2i)$ and $(i,M+2i+1)$ fault.
\item \textbf{Single internal line failure:} Single internal line in
  Area~$i$ faults. Without loss of generality, we consider only
  even line,  that is, line $(i,M+2i)$ faults.
\item \textbf{Tie line and internal line failure:} Internal line in
  Area~$i$, as well as the tie line connected to the corresponding
  demand, fault. Without loss of generality, we consider only an
  even line and its corresponding tie line, that is, $(i,M+2i)$ and
  $(M+(2i-1\mod 2M), M+2i)$ fault.
\end{itemize}
It is important to notice that these failures cover the possible types of
geographical failures over the ring (some combinations of these
failures can also occur if the failure radius is large enough). In
this section, we will not consider sporadic failure events (that is, in
which the outaged lines are not geographically-correlated), or initial
events that partition the graph such that there is a component  whose
total demand is not equal to its total generation.

We next compare between an area failure and parallel lines
failure. As for area failure, it is
easy to see that the same power flow solution still holds, and
therefore, the power flow along all operating lines does not
change.  The loss of demand is only $2$ units and the resulting yield
is $\frac{M-1}{M}$.
On the other hand, upon parallel lines failure, the entire generation of
node $i$ must be transmitted on the only operating lines connected to
node $i$: $(i,M+2i+1)$ and $(i,M+2i+1)'$. Since the lines are
parallel, by symmetry, each of these lines transmits a power of
$1$ unit (cf. Observation~\ref{lem:samesum}).
In addition, by (\ref{eqn:flow1}),  the tie line connecting area $i$ to area $i+1$
carries power of $1$ unit since node $2i+1$'s demand is only 1.
Now focus on Area $i+1$. The total amount of incoming generation to
this area is  $3$ units ($2$ units generated in node $i+1$ and one
along the tie line in
Area $i$), while the total demand is $2$ units. This immediately
implies that the tie line to Area $i+2$ carries one unit of power. As
for the flows on the lines within that area, one can verify that a valid
(and therefore the only) solution of
(\ref{eqn:flow1})--(\ref{eqn:flow2})
has no power flow on lines $(i+1,M+2i+2)$ and  $(i+1,M+2i+2)'$ and one unit
of power flow along  $(i+1,M+2i+3)$ and
$(i+1,M+2i+3)'$. This flow assignment is identical to the
assignment of Area $i$, and therefore, by inductive arguments, it is valid for all
the areas in the ring. Moreover, there are three phase values in the
solution: one
for all generation nodes, one for all even demand nodes, and one for
all odd demand nodes.

Based on the power capacity of the lines $u_{ij}$, the moving average
parameter $\alpha$ of the cascading failure model, and the FoS of the
entire system, we can derive the following results (all proofs are in the appendix):

\begin{lemma}
\label{thm:arbitrarilyfar}
Consecutive failures in a cascade may happen within an arbitrarily
long distance of each other.
\end{lemma}

We note that Lemma~\ref{thm:arbitrarilyfar} captures an important
difference between our model and previously-suggested models, that
assume that power grid failures propagate in an epidemic-like manner.
 While, under these models, a line
failure causes only adjacent node/line (or a line with small distance) to
fault, our model captures situations in which the cascade \emph{``skips''}'
large distances within a single iteration. As we will discuss in
Section~\ref{sec:SanDiego}, this was indeed the case in a recent real-life
cascade causing a major blackout in California.

\begin{lemma}
\label{thm:arbitrarilylarge}
A failure of $o(1)$ of the lines may cause an outage of a constant
fraction of the lines, within one iteration.
\end{lemma}

The following two lemmas show that the failures do not always behave
monotonically:

\begin{lemma}
\label{thm:nonmonotone1}
 An initial failure event of some set of lines $A$ may result in a
 lower yield than a failure
event, whose initial set of faulted lines is a superset of $A$.
\end{lemma}

\begin{lemma}
\label{thm:nonmonotone2}
 A network $\mathcal{G}_1$, whose topology is a subgraph of the
 topology of another network $\mathcal{G}_2$, may obtain
higher yield.
\end{lemma}

We note that in practice, if the failures are geographically
correlated, non-monotone situations, as described in
Lemmas~\ref{thm:nonmonotone1} and~\ref{thm:nonmonotone2}, rarely
happen. Thus, in the rest of the paper (and specifically, in the
algorithm described in Section~\ref{sec:algos}), we assume only a
monotone behavior of the failures.


Up until now, we dealt with very
specific type of failures, which intuitively \emph{breaks the ring into a
chain.} In general, such failures are easier to analyze since
(\ref{eqn:flow2}) does not form contraints in a cyclic manner
(that is, one can assign the phase value of the nodes based only on
the flows and the phase value of the nodes that precedes it). More
formally, this extra difficulty is captured by
Observation~\ref{lem:samesum},
and looking at the different paths between each pair of nodes.
When the ring breaks into a chain, all these paths
follow one ``direction''. On the other hand, when the ring does
not break, there are both clockwise and counter-clockwise paths that
need to be considered.   We demonstrate this difference by comparing a
single internal line failure of line $(0,M)$ (that does not break the
ring into chain) with a tie line and internal line failure of lines
$(0,M)$ and $(3M-1,M)$ (which breaks the ring into chain).

In the tie line and internal line failure, one can simply assign a
power flow of $1$ unit along the parallel line $(0,M)'$, where the phase
difference between node $0$ and $M$ is $1$ (to meet the constraint of
(\ref{eqn:flow2})); the rest of the flows and phases remain
unchanged. However, this solution is not valid in case of a single internal
line failure, since there is a phase difference between node
$M$ and $3M-1$, and therefore, it is impossible that no flow
traverses the operating line between them. We note also that in the
odd and even lines failure (which does not break the chain either),
the flows on the lines parallel to the failures increase to $1$ by
changing solely the generator phase. Since the demand nodes'
phases do not change, this failure is localized within that single
area.

We next show that the power flow values induced by the single line failure
change across the entire graph and depends on the value of
$M$:
\begin{lemma}
\label{lemma:singlefailure}
Consider an $M$-ring, in which line $(0,M)$ failed. Then, after the
first iteration: \\
(i) The flow along line $(0,M)'$ is $2M/(2M+0.5)$.  \\
(ii) The flow along all other even lines is $M/(2M+0.5)$. \\
(iii) The flow along all odd lines is $1-M/(2M+0.5)$. \\
 (iv) The flow along all tie lines is $1-2M/(2M+0.5)$.
\end{lemma}


\begin{corollary}
\label{cor:ninth}
An $M$-ring requires power capacity of $1/9$ on its tie lines and a
power capacity of $1$ on its internal lines to withstand any
failure of one line.
\end{corollary}

Interestingly, one can see that as $M$ gets larger, a single internal
line failure has the same effect as the corresponding tie line and
internal line failure. This stems from the fact that the closed-loop
effects, initially distinguishing between the failures, are fading
away as the ring gets larger.

Finally, the next lemma shows that, unlike the examples we presented
on the ring, cascades can be made arbitrarily long (in time). The
lemma uses another topology which is depicted in
Figure~\ref{fig:parallel}. 

\begin{lemma}
\label{lem:long}
The length of the cascade (the number of iterations until the system stabilizes) can be
arbitrarily large.
\end{lemma}

To conclude, using simple examples we highlight the difficulties and differences between
prior models used to analyze the power grid and the models we use.
In the rest of the paper, we will investigate how these models behave
in real-life power grid and geographically-correlated failures.



%% file: algos.tex
\section{Power Grid Resilience}

\label{sec:res}
\subsection{Parameters Set-up}
\label{sec:capacities}
We note that in the cascading failure model, the power capacities $u_{ij}$  of the lines  are given a-priori. In practice, however,  these capacities are hard to obtain and are usually estimated based on the actual operation of the power grid.  In this paper, we take the $N-k$ contingency analysis approach~\cite{Dan} in order to estimate the power capacities. Namely, we set the capacities so that the network is resilient to failure of any set of $k$ out of the $N$ lines. In addition, we consider over-provisioning of lines capacity by a constant fraction of the required capacity of each line. This over-provisioning parameter, denoted by $K$, is often referred to as the \emph{Factor of Safety} (FoS) of the grid.

Specifically, we focus on the following two cases.
\begin{itemize}
\item \textbf{$N$-resilient grids} (that is, $k=0$). In this case, we solve (\ref{eqn:flow1})--(\ref{eqn:flow2}) for the original grid graph (without failures) and set the power capacity to $u_{ij} = K\cdot f_{ij}$, where $K \geq 1$.
\item \textbf{$(N-1)$-resilient grids} (that is, $k=1$). In this case, we solve (\ref{eqn:flow1})-- (\ref{eqn:flow2}) for $N$ graphs, each resulting from a single line failure event. The power capacity is set to $u_{ij} = K\cdot\max_r f_{ij}^r$, where $f_{ij}^r$ is the flow assigned to line $(i,j)$ when considering the $r^{\mbox{th}}$ failure event. 
\end{itemize}

It is worth mentioning that the real power grid is usually assumed to be at least $N$-resilient with $K \approx 1.2$~\cite{DobsonPersonal}. On the other hand, some data show that certain lines (or, more generally, \emph{paths}) are more resilient than others. For example, a historical transmission paths data found in~\cite{WECC_hist} shows that some transmission paths have power capacities which are $1.1$ times their normal flow, while others have an FoS larger than $2$. Nevertheless, the average FoS is indeed around $1.2$. In addition, utility companies usually guarantee  that their grid is at least $(N{-}1)$-resilient~\cite{Dan}\footnote{We note that early reports on the recent San Diego blackout indicate that this was not the case.}. Therefore, by setting the power capacities parameters, we examine in this paper both $N$- and $(N{-}1)$-resilient grids with different FoS values $K$. Most of our numerical results are presented for a grid with FoS $K=1.2$.

\subsection{Identification of Vulnerable Locations} \label{sec:algos}

We consider a circular and deterministic \emph{failure model}, where all lines and nodes within a radius $r$ of the failure's epicenter are removed from the graph (this includes lines that pass through the affected area). In addition, we assume \emph{monotonicity} of failures: if the initial set of faulted lines due to event $A$ is a subset of the faulted lines due to event $B$, then the yield of $A$ is at least that of $B$. We note that in the general case, this property does not hold (see Lemma~\ref{thm:nonmonotone1}). However, we observed that such events rarely happen in real power grid systems, and  when they do, it is only when both events have a marginal effect. Since our goal is to identify the most vulnerable locations in a real power grid, this is a valid assumption for practical purposes.

To identify the candidates for the most vulnerable locations, we use computational geometric methods developed in \cite{infocom11} for identifying the vulnerable locations in fiber-optic networks. For each line, we define an $r$-\emph{hippodrome}, which captures all points in the plane $\reals^2$ whose distance from the line is at most the failure radius $r$. We focus on the \emph{arrangement of hippodromes}, which is the subdivision of the plane into vertices, arcs, and faces. The vertices are the intersection points of the hippodromes, the
arcs are either maximally connected circular arcs or straight
line segments of the boundaries of hippodromes that occur
between the vertices, and faces are maximally connected
regions bounded by arcs. 

Once the vertices of the arrangements are identified, we treat each vertex $v$ as a \emph{candidate} for a failure epicenter and denote by $\mathcal{L}(v)$ the set of lines within radius $r$ of $v$ ($\mathcal{L}(v)$ can be easily found). We then use the Cascading Failure Model, described in Section \ref{sec:power_mod},
 with $\mathcal{L}(v)$ as the set of lines that initially fault. Naturally, the process of checking all candidates (each with a different initial failure event) can be easily parallelized.

Arrangements are well-established concept in computational geometry, and it can be easily shown that in order to find the vulnerable locations, it is sufficient to consider only the vertices of the arrangements. In particular, for any point $p\in\reals^2$, there is a vertex $v$ such that $\mathcal{L}(p)\subseteq \mathcal{L}(v)$.  Notice that computing arrangements is quadratic in the number of lines. Thus, we parallelized this computation as well by partitioning the graph into several sections (with small number of lines) and finding vertices of the arrangements in each section. To ensure that no vertices are lost in the border between two sections, the sections have a $2r$ overlap.

%% file: data.tex
\section{Power Grid Data}
\label{sec:data}
We use \emph{real power grid data} of the western US taken from the Platts Geographic Information System (GIS) \cite{GIS}. This includes the information about the transmission lines, power substations, power plants, and the population at each geographic location. Since, in GIS each transmission line is defined as a link between two power substation, substations are used as nodes in our graph.
In order not to expose the vulnerability of the real grid, we used a \emph{part} of the Western Interconnect system which does not include the Canada and Mexico sections. On the other hand, we attached to the grid the Texas, Oklahoma, Kansas, Nebraska, and the Dakotas' grids, which are not part of the Western Interconnect. The resulting graph (see
Figure~\ref{fig:network}) has $14{,}968$ nodes (substations) and $19{,}513$ lines. Moreover, it has $1{,}920$ power stations which were merged with the substations as described below. We note that there is a small number of very dense areas (e.g., the Los Angeles area), while the rest of the grid is very sparse. This structure can be seen in many other typical power grids, such as the US Eastern Interconnect as well as European systems. Furthermore, recent research on topological models for power grid systems show similar results \cite{vermont}. Thus, our results will probably carry over to other grids.

 We performed the following processing steps for this graph.
\par \noindent \textbf{Coordinate transformation:} The coordinates of the substations in the GIS system are given by their \emph{longitude and latitude}. We transformed these to \emph{planar} $(x, y)$ coordinates, using the \emph{great-circle distance} method.  

\par \noindent \textbf{Connectivity check:} We identified the connected components of the raw GIS graph, which consists of one large connected component and few small ones. Moreover, we identified some substations that appear in the transmission lines data but are absent in the substations data. The number of these substations is small, and therefore, after manual inspection, they were either eliminated or merged with other nearby substations  (see also the next step).

\par \noindent \textbf{Nodes merging and lines elimination:} For each node outside the large component, we found the closest node within the large component. If the distance between them was below a given threshold ($10$ Km), we merged the two nodes. Then, the remaining disconnected nodes were inspected manually and were either eliminated or merged.
At the end of this process, we obtained a fully-connected graph (that is, a single connected component) with $13{,}992$ nodes and $18{,}681$ lines.

\par \noindent \textbf{Identifying demands and supplies:}
Demands were associated with the closest node to each populated area (i.e., zip-code) and were set to be proportional to the population size (the normalization factor is computed by comparing the total population and total generation output of the entire grid). Supplies were associated with closest node to  each power plant (the generation capacity of the power plants is taken from the GIS). Then, for each node, we computed the difference between its total corresponding supplies and total corresponding demands. Thus, all nodes were characterized by a real number: positive for a supply node,  negative  for a demand node, and zero for a neutral node. The resulting categorization appears in Figure~\ref{fig:network}. Overall, $1{,}117$ nodes were classified as generators (supplies), $5{,}591$ as loads (demands), and $7{,}284$ as neutral. Most of the neutral nodes are closely connected to each other and to one of the non-neutral nodes, thus drawing the power/demand from them.

\par \noindent \textbf{Determining the system parameters:} The GIS does not provide the power capacities of the transmission lines, nor their reactance. However, these parameters are needed for the power flow and cascading failure models.  The reactance of a line depends on its physical properties (such as its material) and there is a linear relation between its length and reactance: the longer the line is, the larger its reactance. Thus, we assumed that all lines have the same physical properties (other than length) and used the length to determine the reactance. It is important to notice that the flow part of the solution of (\ref{eqn:flow1})--(\ref{eqn:flow2}) is scale invariant to the reactance (that is, multiplying the reactance of all lines by the same factor does not change the values of the flows). Thus, we simply use the length of the line as its reactance.  
Regarding the power capacities, we take the approach described in Section \ref{sec:capacities}. In particular, we use $N-k$ contingency analysis, with $k \in\{ 0, 1 \}$ and different FoS values $K$.

%% file: results.tex
\section{Numerical Results} \label{sec:results}

We identified the potential failure locations using the algorithm described in Section \ref{sec:algos} implemented in MATLAB. We present results for $r  = 50$  kilometers, which is small enough to capture realistic scenarios~\cite{EMP,EMP_flare}, while it is large enough to generate a cascading failure in most cases (the results for other values of $r$ are omitted, due to space constraints). For $r=50$, the algorithm identified $61{,}327$ potential failure locations. The identification of these locations was done on an eight-core server in less than $24$ hours.

For each failure location $v$, we performed the simulation of the Cascading Failure Model, presented in Section \ref{sec:power_mod}, assuming that all lines in $\mathcal{L}(v)$  fail. The simulation was performed using a program that efficiently solves very large systems of linear equations, using  CPLEX \cite{CPLEX} and Gurobi \cite{gurobi} optimization tools.

To assess the severity of a cascading failure, we use the following four metrics, which are measured in the end of the cascade:
\textbf{The yield}, as defined in (\ref{eqn:tp});
\textbf{the total number of outaged lines}, which indicates the time it takes to recover the grid after the cascade: the larger the number of outaged lines, the longer is the actual time of the corresponding blackout;
\textbf{the number of connected components}; and
\textbf{the number of rounds until stability}.


\subsection{$\mathbf{N}$-Resilience Experiments}
The first set of experiments was performed after using the $N$ contingency analysis to set the capacities of the network, with FoS $K=1.2$ and $K=2$.

\begin{figure}
\centering
\subfigure[San Diego area failure event.]{\epsfig{figure=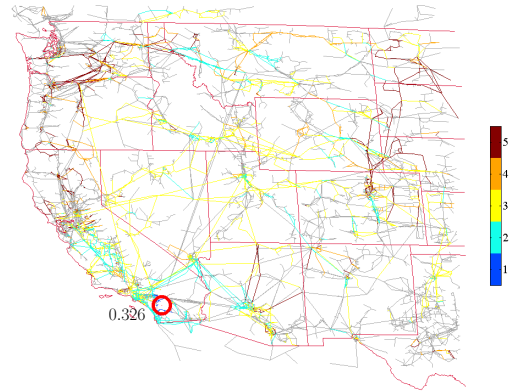, width=0.37\textwidth}} \\
\subfigure[San Francisco area failure event.]{\epsfig{figure=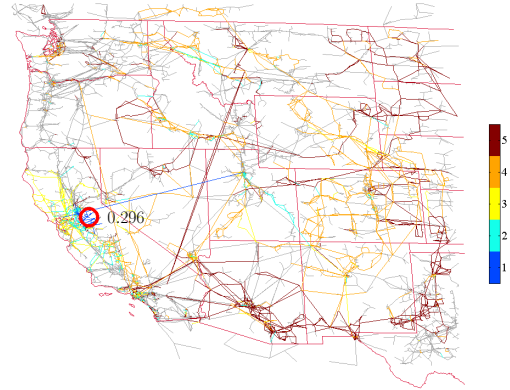, width=0.37\textwidth}} \\
\subfigure[Idaho-Montana-Wyoming border failure event.]{\epsfig{figure=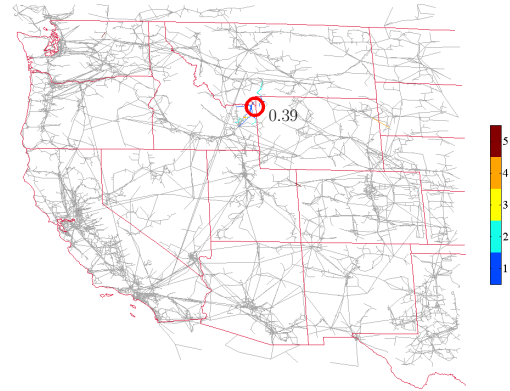, width=0.37\textwidth}}
\caption{Illustration of cascading failures over $5$ rounds for $N$-resilient grid with FoS $K=1.2$, where the initial failure locations are in the (a) Los Angeles area, (b) San Francisco area, and (c) Idaho-Montana-Wyoming border. The final yields are 0.326, 0.296, and 0.39, respectively. The colors represent the rounds in which the lines faulted. \label{fig:cascades_5_1}}
\end{figure}

First, we plot specific failures to show how they evolved during the first \emph{five} rounds of the cascade. Figure \ref{fig:cascades_5_1} shows three  failure events for FoS $K=1.2$: Two in California, both leading to severe blackouts, and another one around the Idaho-Montana-Wyoming border, which had a less severe effect. The round-by-round maximum overload (that is, $\max_{ij} f_{ij}/u_{ij}$) for these failures and $K=1.2$ is shown in Table~\ref{fig:cascades_data_5}. The same failures for $K=2$ are shown in Figure \ref{fig:cascades_5_2} in the appendix. As expected, higher FoS usually leads to less severe blackout effect. Interestingly, the Idaho-Montana-Wyoming border failure with FoS $K=1.2$ leads to low yield (0.39), although the development of the failure is very slow---after 5 rounds only few lines were faulted. However, the same event with $K=2$ leads to near-unity yield. We note that this suggests that the assumption that $K=1.2$ for all lines is quite pessimistic, as also can be seen from the actual data (see Section \ref{sec:capacities} for more details).


\begin{table}
\centering
\epsfig{figure=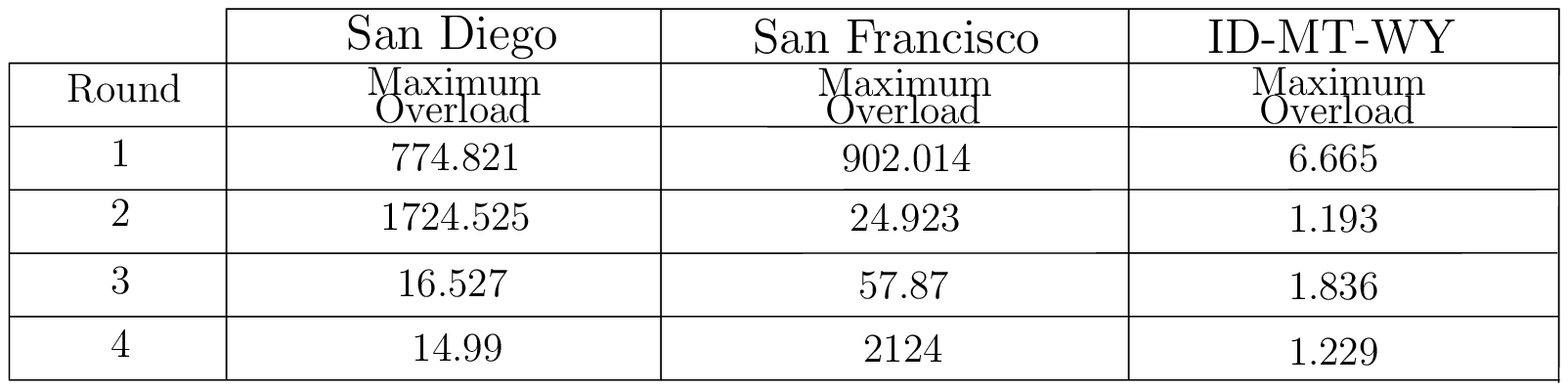, width=0.49\textwidth}
\caption{Round-by-round maximum line overload $\max_{ij} f_{ij}/u_{ij}$  for the cascading failures shown in Figure \ref{fig:cascades_5_1}.  \label{fig:cascades_data_5}}
\end{table}

\begin{figure}
\centering
\epsfig{figure=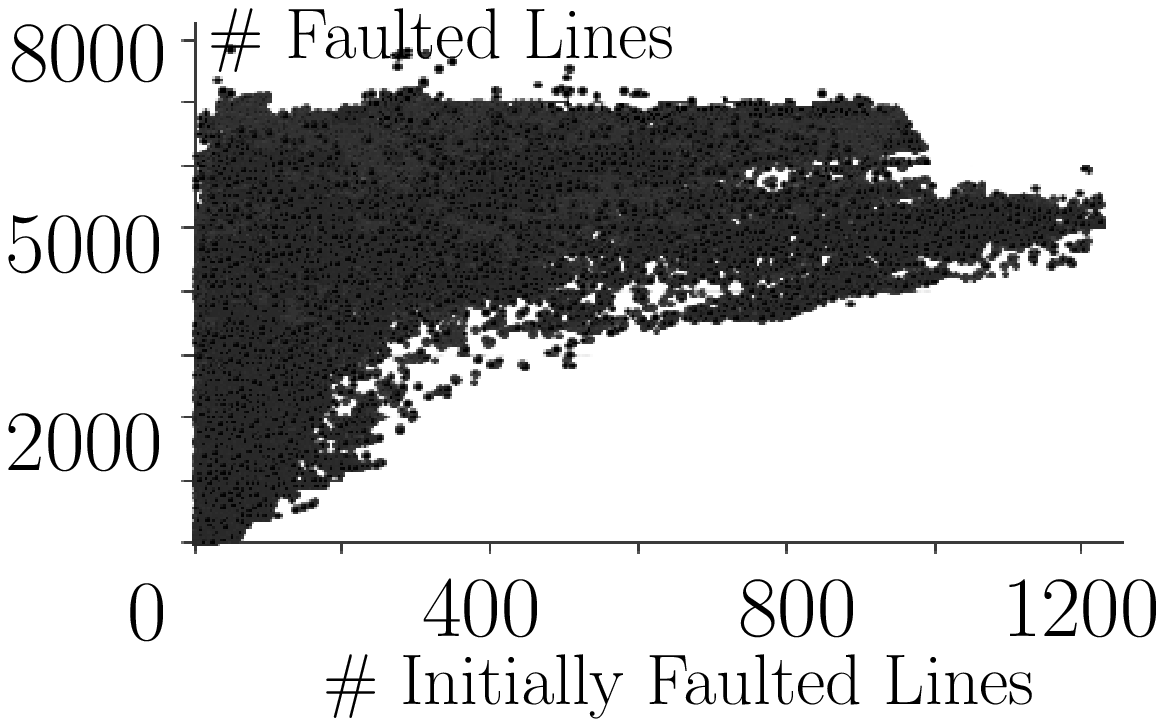, width=0.23\textwidth}
\epsfig{figure=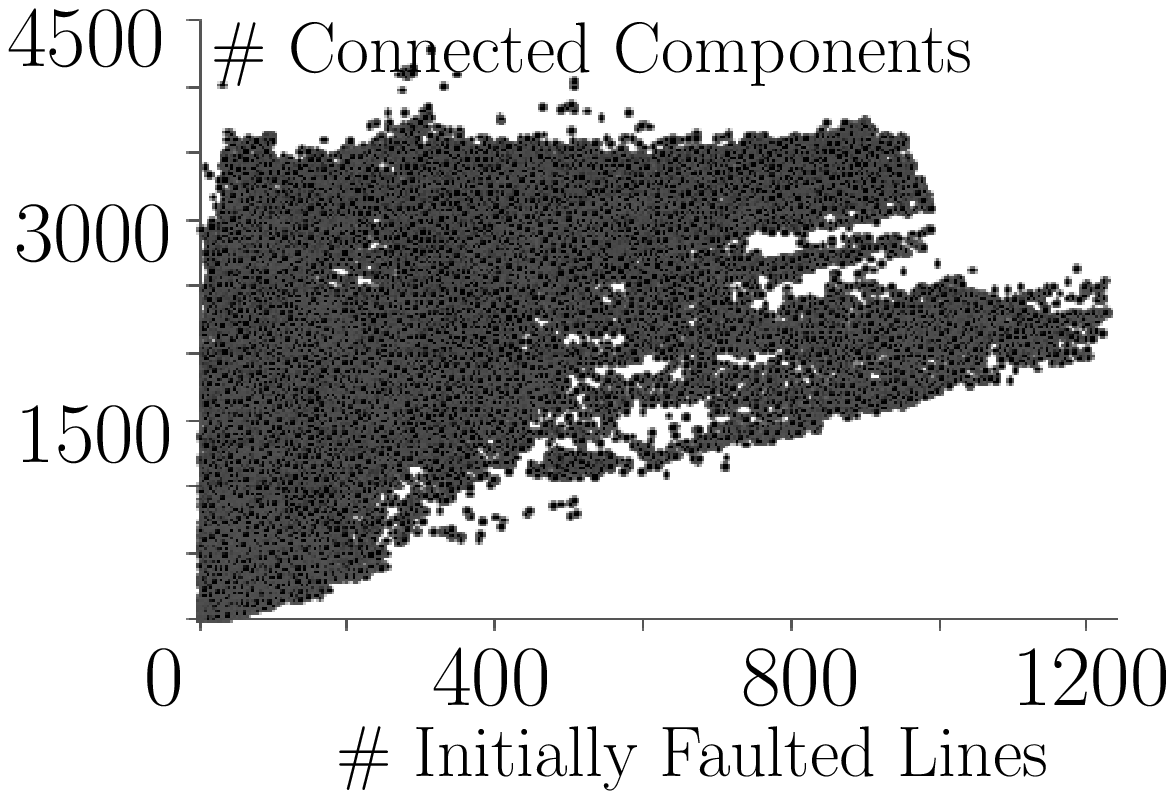, width=0.23\textwidth}
\caption{The effects of the number of initially faulted lines on the total number of faulted lines (left) and the number of components (right),  after $5$ rounds of cascade (FoS  $K=1.2$\label{fig:params_5}).}
\end{figure}

Scatter graphs for different metrics after $5$ rounds and with FoS $K=1.2$ are shown in Figure \ref{fig:params_5}.
It can be seen an increase in the initial number of faulted lines leads to an increase in the total number of faulted lines at the end of the fifth round: if $400$, $800$,  $1{,}200$  lines initially faulted, at least $2{,}847$, $3{,}600$, $4,669$  are faulted at the end, respectively.
Furthermore, an
increase in the initial number of faulted lines leads also to an increase in the number of connected components: if $400$, $800$,  $1{,}200$  lines initially faulted, the number of components is at least $696$, $1{,}382$, $1{,}973$, respectively.
Similar results for $K=2$ are shown in Figures \ref{fig:cascades_5_2} and \ref{fig:params_5_2} in the appendix. These results clearly show that in this case the grid is much more resilient to cascading failure. 

\begin{figure}
\centering
\subfigure[The yield as a function of the number of initially faulted lines (left) and the yield as a function of total number of faulted lines (right) .\label{fig:params_40_yield_lines}]{
\epsfig{figure=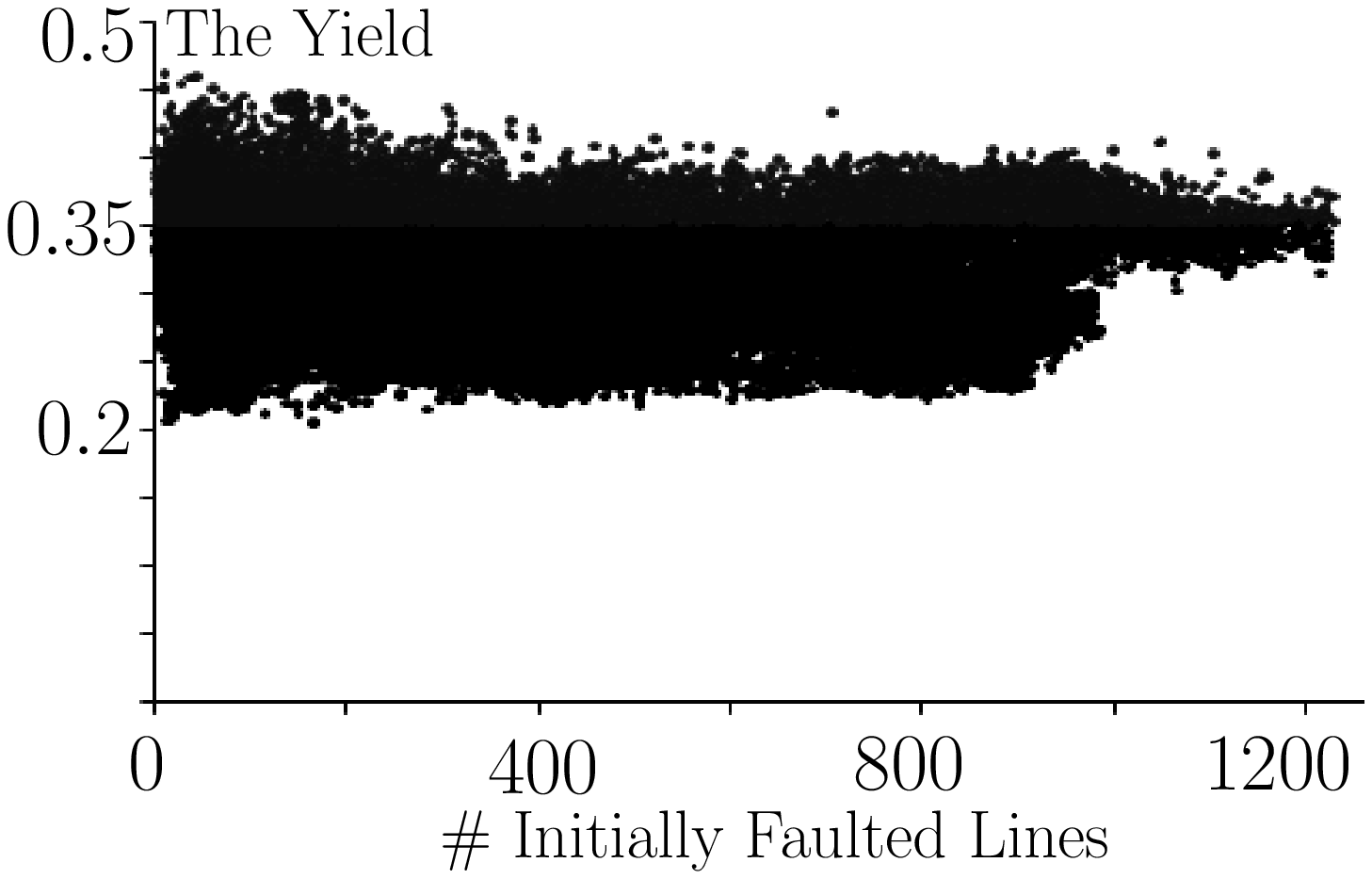, width=0.23\textwidth}
\epsfig{figure=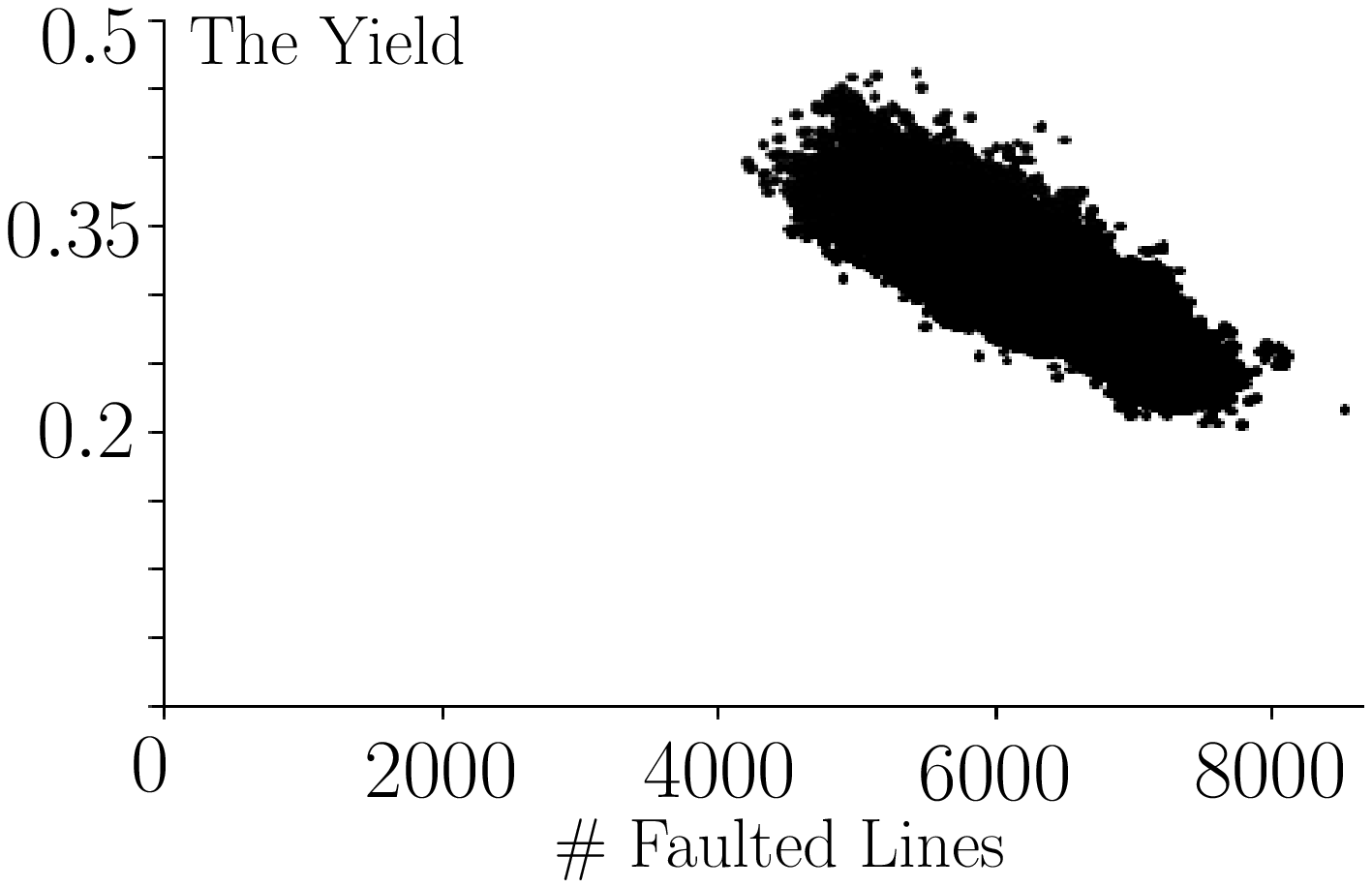, width=0.23\textwidth}
}
\subfigure[The number of rounds as a function of the number of initially faulted lines (left) and the yield as a function of the number of rounds (right).\label{fig:params_40_yield_rounds}]{
\epsfig{figure=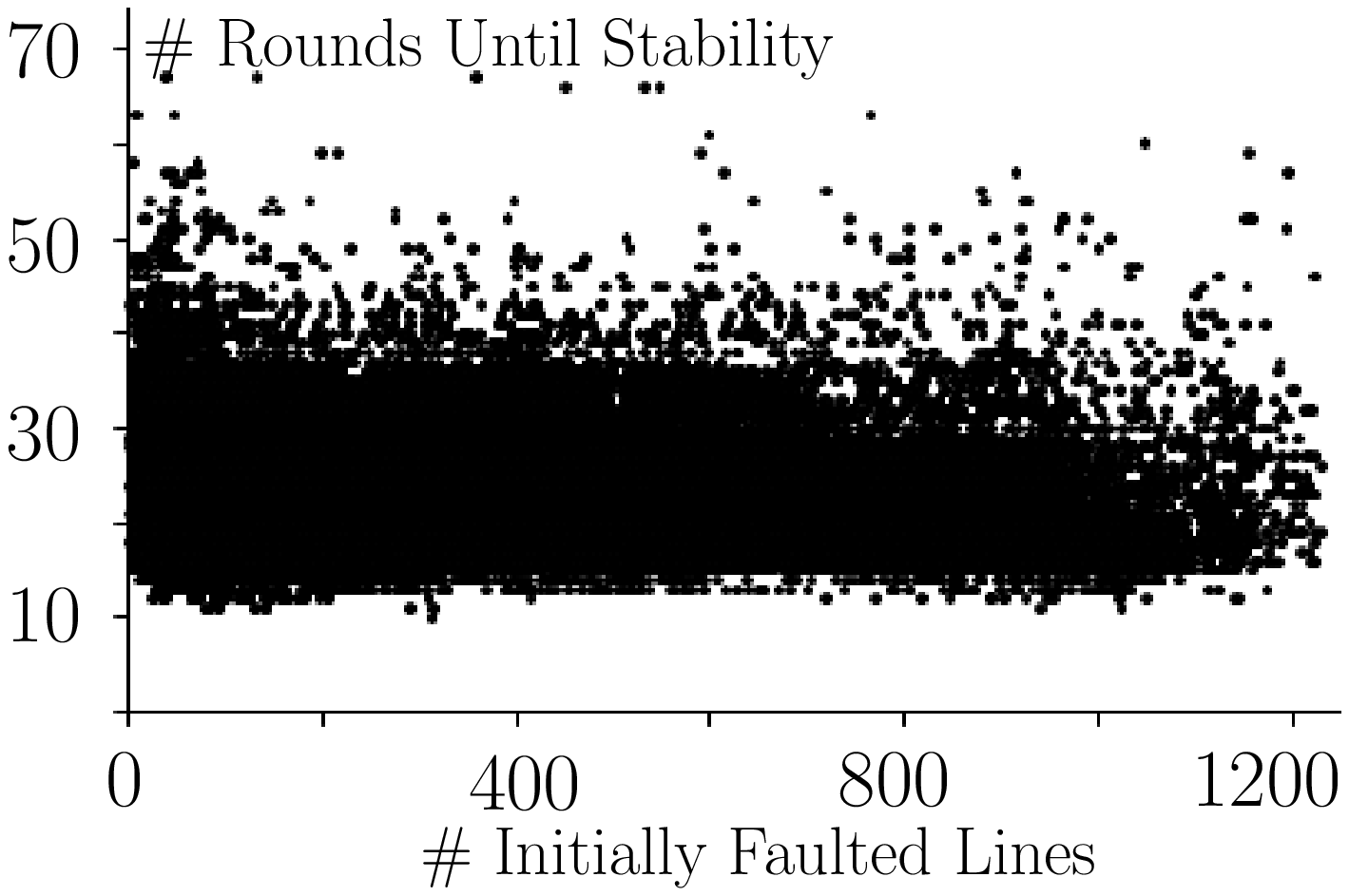, width=0.23\textwidth}
\epsfig{figure=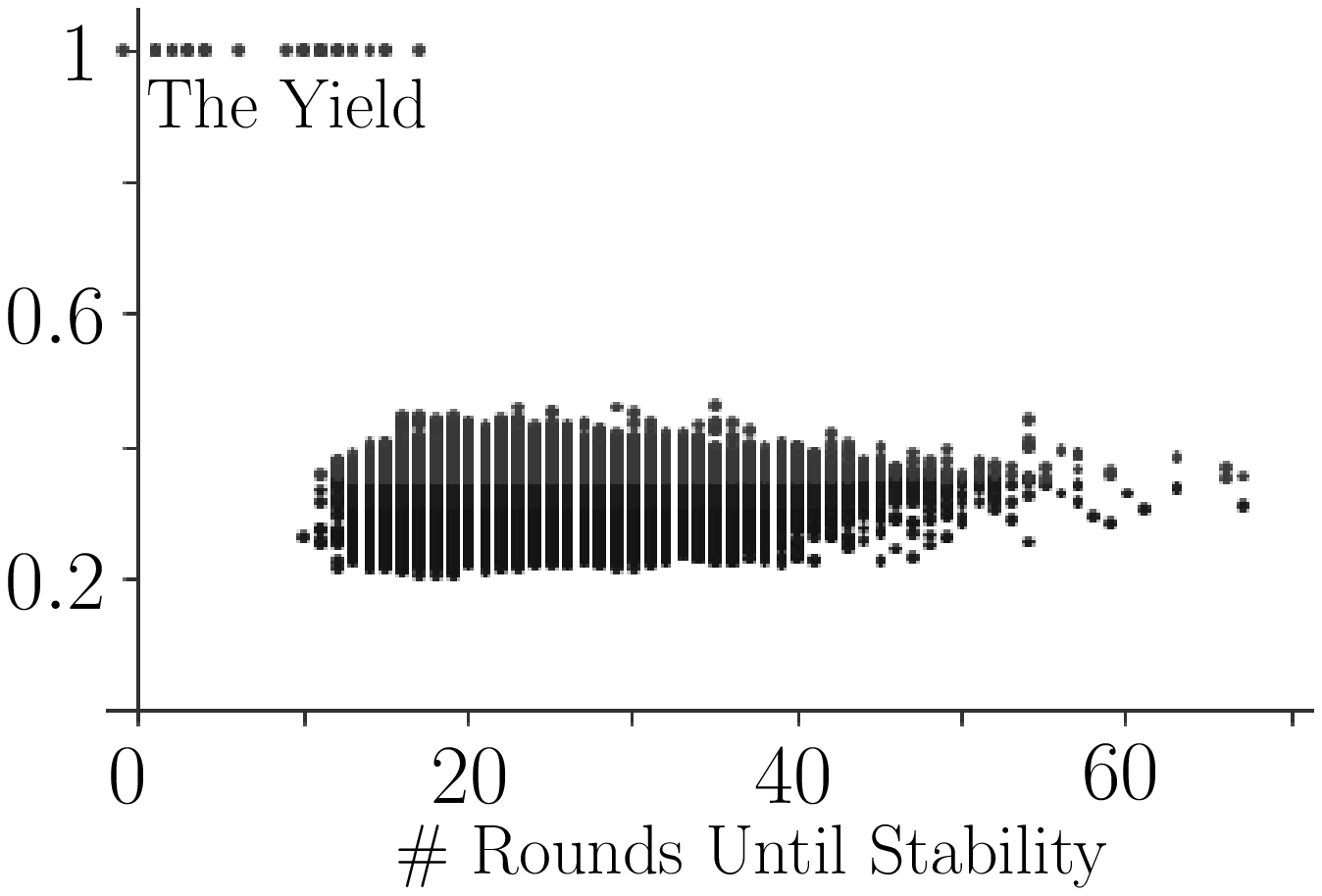, width=0.23\textwidth}
}
\caption{Comparison between different performance metric at stability (FoS $K=1.2$). \label{fig:params_40}}
\end{figure}

Next, we analyze the severity of cascading failures once \emph{stability} is reached. Namely, when no more line failures occur. The results for FoS $K=1.2$ are shown in Figure \ref{fig:params_40}. In this case, the vast majority of the failures resulted in yield in the range of  $0.2$--$0.46$.

Figure \ref{fig:params_40_yield_lines} focuses on points whose yield is in $[0.2, 0.46]$. The variance of the yield is larger when the initial number of faulted lines is smaller. Moreover, as expected, there is an inverse correlation between
the yield and the total number of faulted lines. 
Figure \ref{fig:params_40_yield_rounds} shows the relation between the number of initially faulted lines and the number of rounds until stability, and in turn, the relation between the latter and the yield. 
We can see a \emph{threshold effect} in the yield: when the number of rounds is small, the yield is around $1$, while when more than $10$ rounds are required, the yield drops to $0.5$ and below. The correlation between the number of rounds and the number of initially faulted lines suggests (somewhat surprisingly) that usually a smaller number of  initially faulted lines leads to a larger number of  rounds until stability.

\begin{figure}
\centering
\subfigure[The yield values at stability.\label{fig:yield_map_unlim}]{
\epsfig{figure=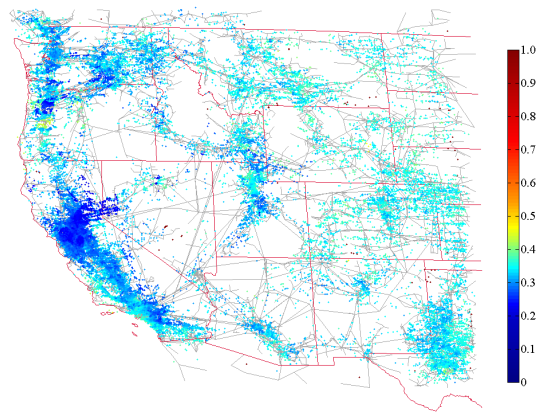, width=0.4\textwidth}
}
\subfigure[The number of rounds until stability.\label{fig:rounds_map_unlim}]{
\epsfig{figure=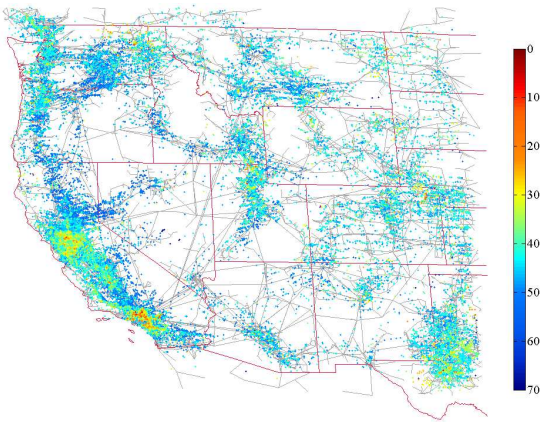, width=0.4\textwidth}
}
\subfigure[The number of faulted lines at stability.\label{fig:rounds_map_unlim}]{
\epsfig{figure=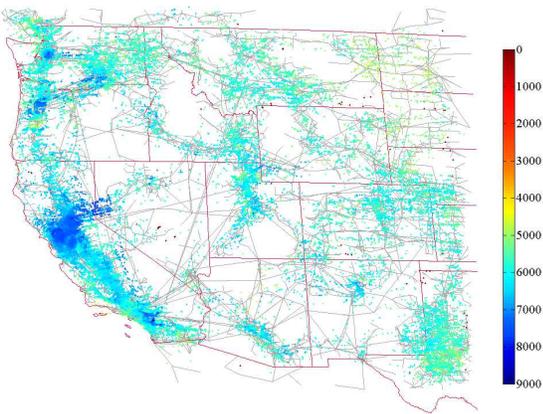, width=0.4\textwidth}
}
\caption{Vulnerability analysis (at stability) of failure locations for $N$-resilient grid with FoS $K=1.2$.  The color of each point (which is a vertex of the arrangement) represents the value corresponding to a cascade whose epicenter is at that point (points that do not appear on the map cause outages that are a subset of the outages caused by a nearby vertex).\label{fig:map_unlim}}
\end{figure}

Finally, Figure \ref{fig:map_unlim} illustrates the yield, the number of rounds until stability, and the number of faulted lines at stability (respectively) by failure location.

\subsection{$\mathbf{(N{-}1)}$-Resilience Experiments}
\begin{figure}
\centering
\epsfig{figure=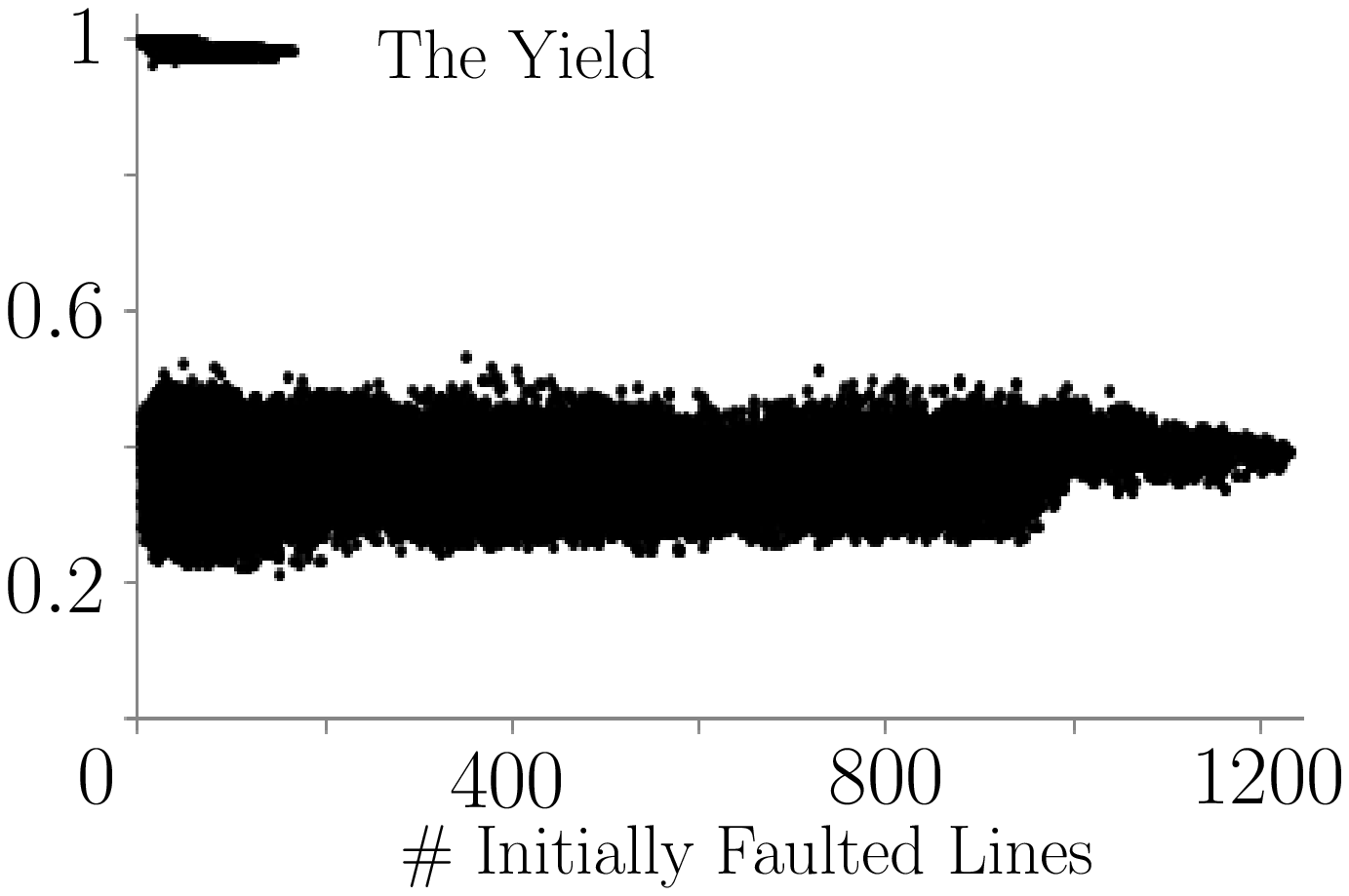, width=0.23\textwidth}
\epsfig{figure=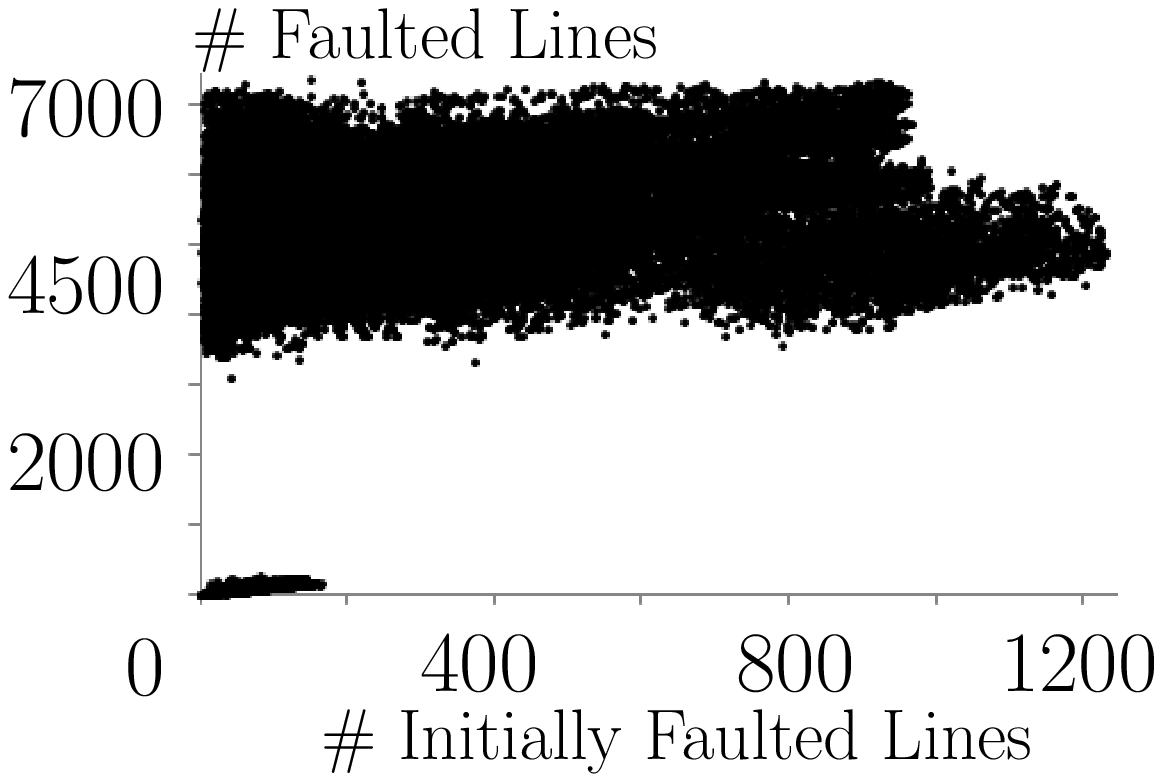, width=0.23\textwidth}
\epsfig{figure=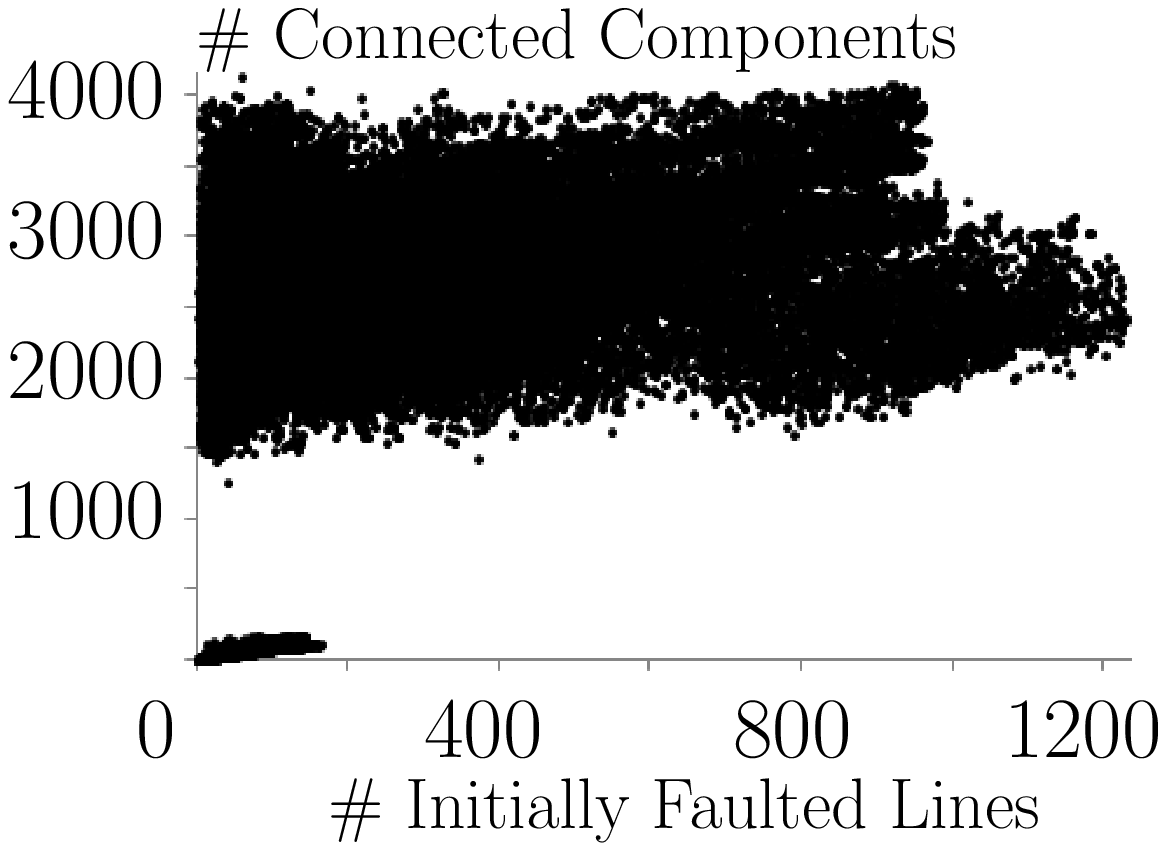, width=0.23\textwidth}
\epsfig{figure=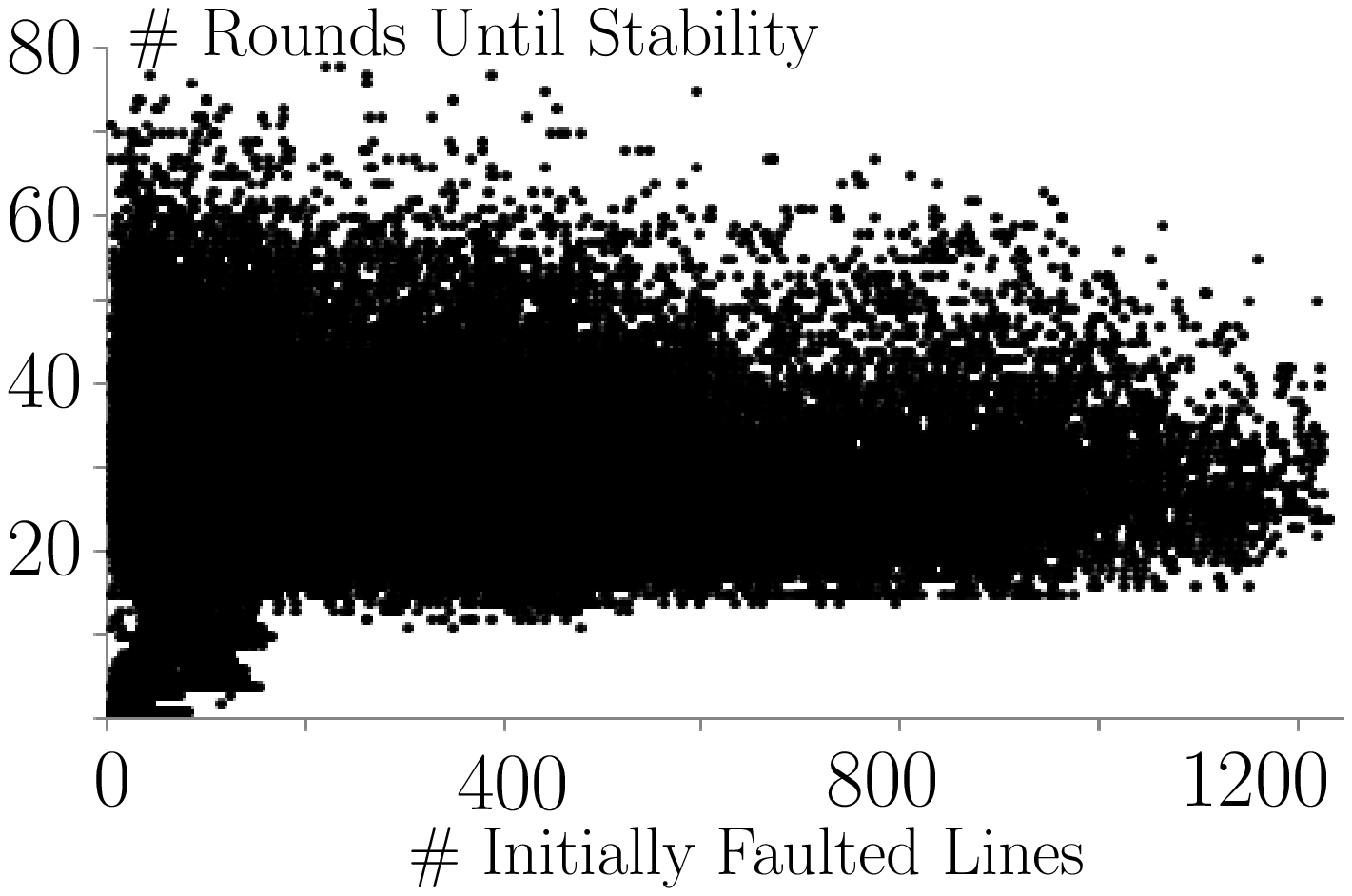, width=0.23\textwidth}
\caption{Comparison between different performance metrics at stability;$(N{-}1)$-resilient grid with FoS $K=1.2$.\label{fig:params_stab_sens}}
\end{figure}

\begin{figure}
\centering
\subfigure[The yield values at stability.\label{fig:yield_map_unlim}]{
\epsfig{figure=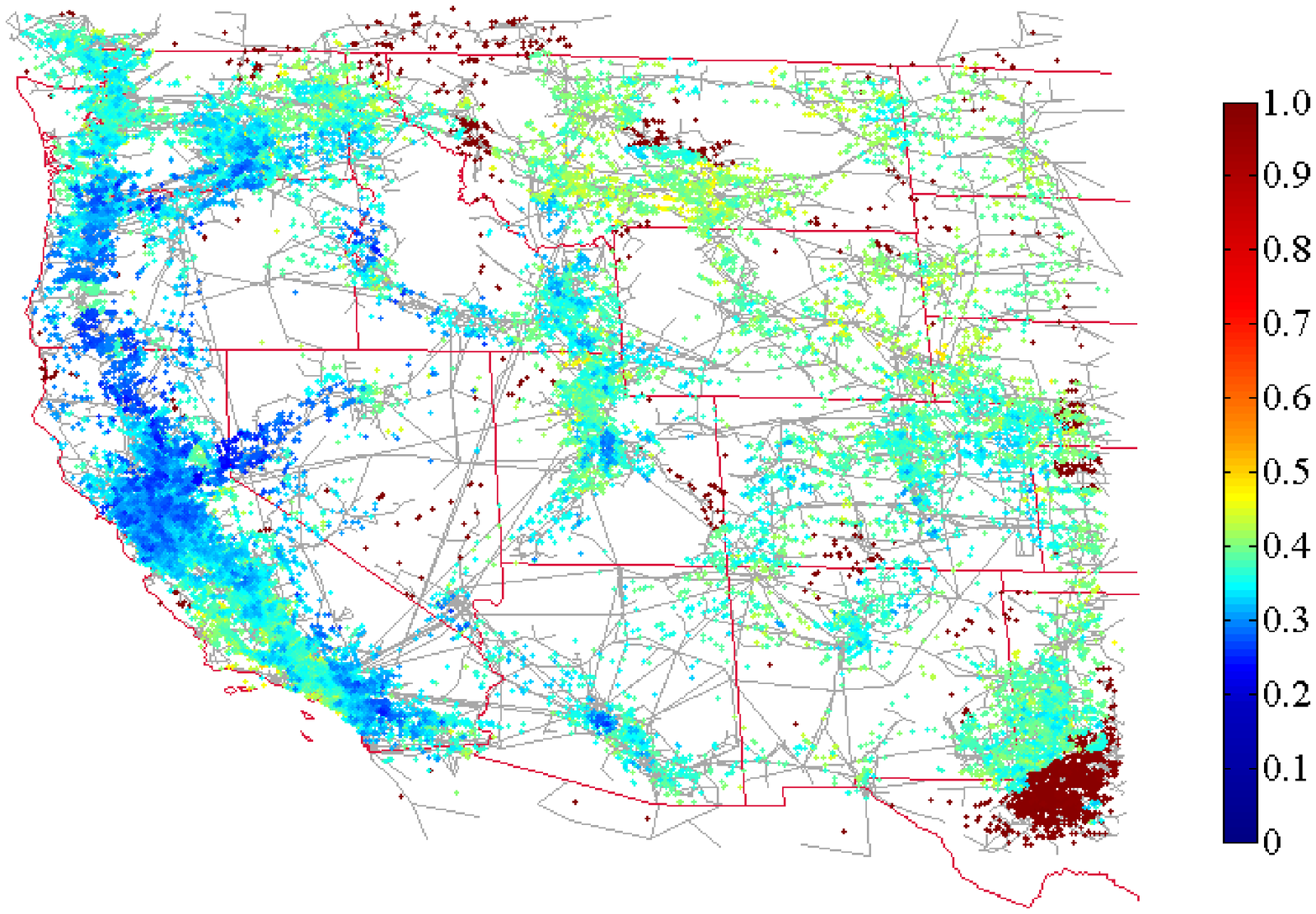, width=0.4\textwidth}
}
\subfigure[The number of rounds until stability.\label{fig:rounds_map_unlim}]{
\epsfig{figure=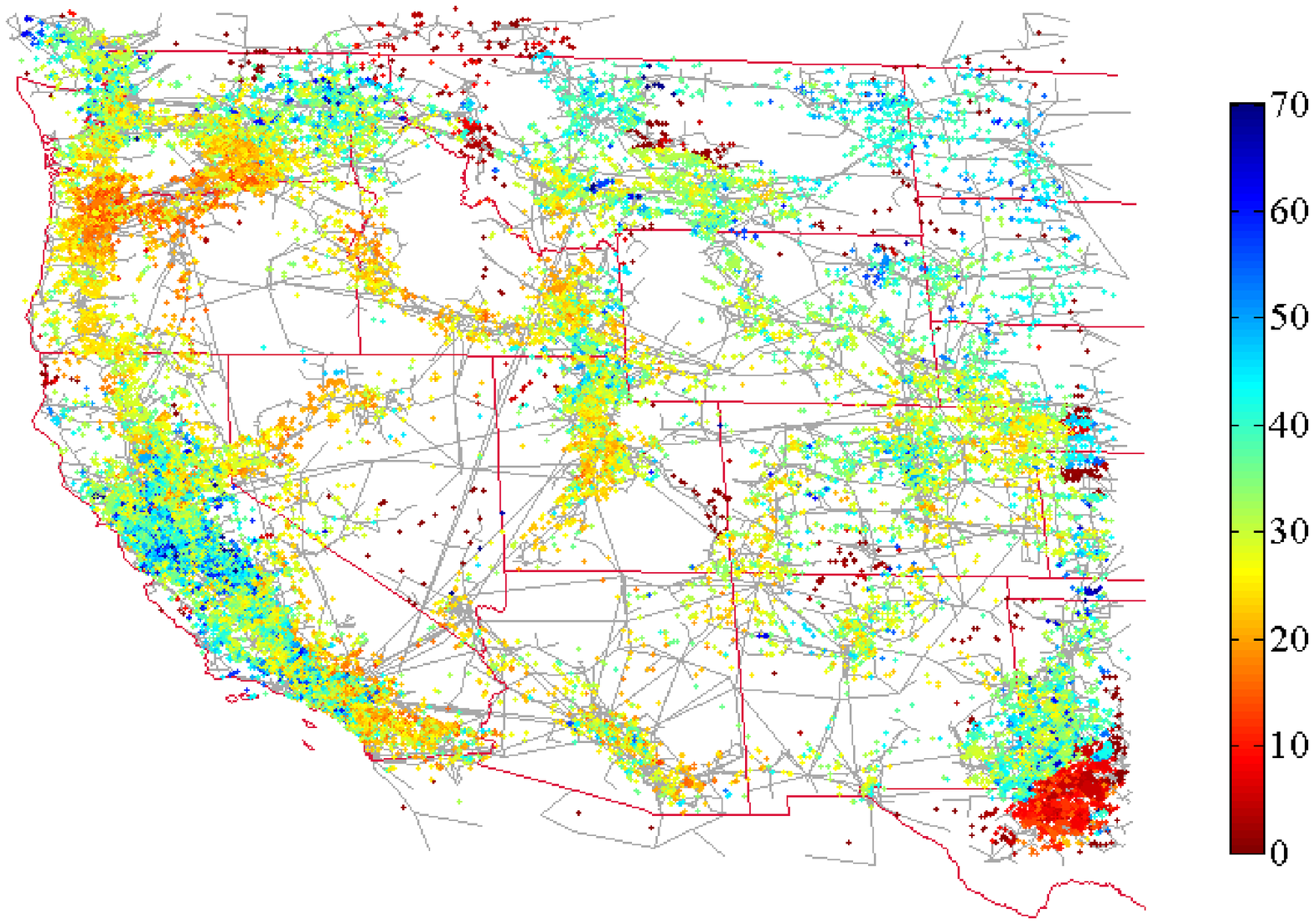, width=0.4\textwidth}
}
\subfigure[The number of faulted lines at stability.\label{fig:rounds_map_unlim}]{
\epsfig{figure=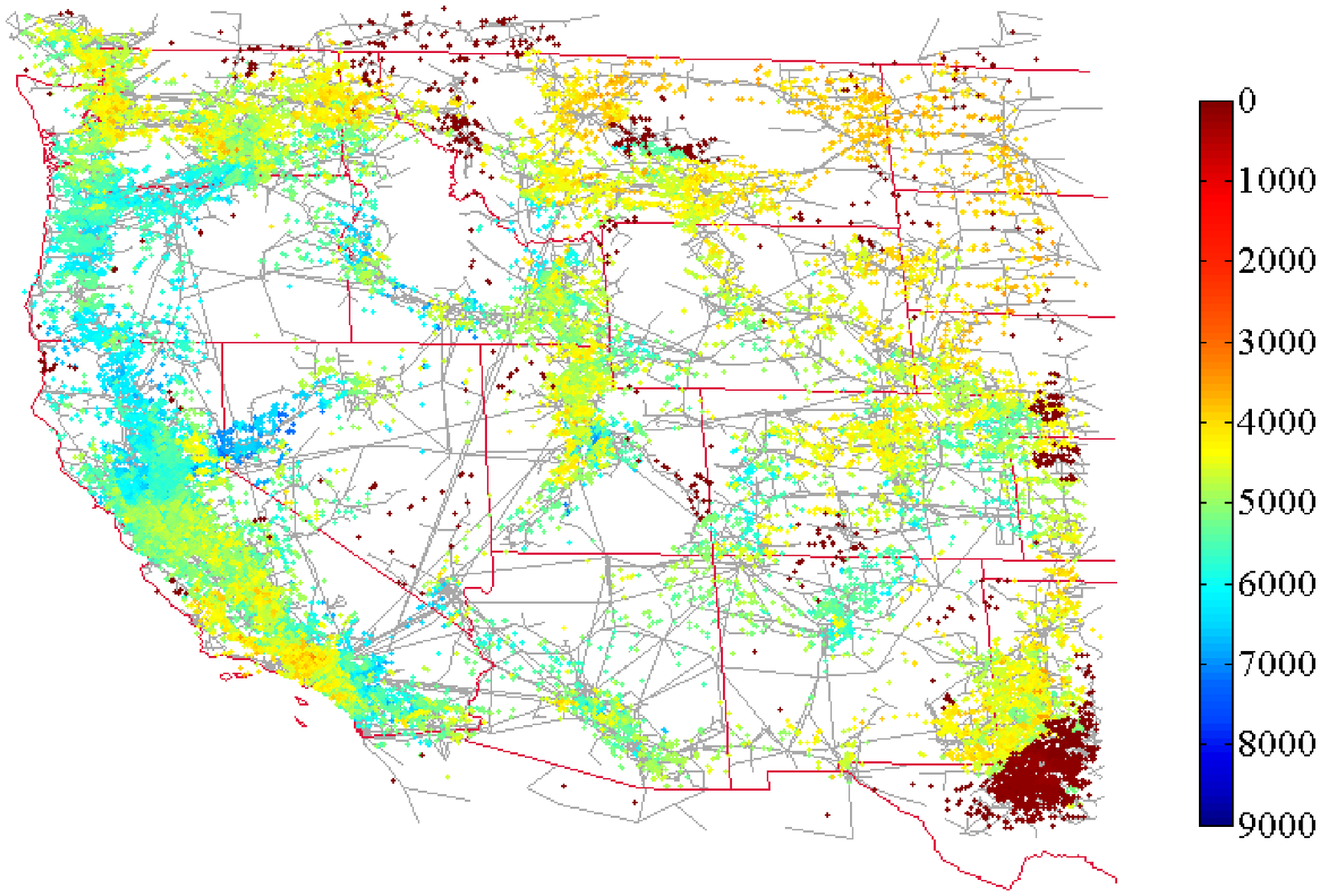, width=0.4\textwidth}
}
\caption{Vulnerability analysis (at stability) of failure locations for an $(N{-}1)$-resilient grid with FoS $K=1.2$.  The color of each point (which is a vertex of the arrangement) represents the value corresponding to a cascade whose epicenter is at that point.
\label{fig:map_unlim_sens}}
\end{figure}

The second set of experiments was performed after using the $N-1$ contingency analysis to set the capacities of the network, with FoS $K=1.2$. The results are presented in Figures 
\ref{fig:params_stab_sens} and \ref{fig:map_unlim_sens}.
Regarding the failure events indicated in Figure~\ref{fig:cascades_5_1}, the corresponding yield values at stability in this case are $0.352$, $0.333$ ,and $0.999$, respectively. The comparison between the results of $N$- and $(N{-}1)$-resilience with the same FoS ($K=1.2$) suggests, as expected, that $(N{-}1)$-resilience helps when the initial event is not significant (such as the Idaho-Montana-Wyoming border event). However, it makes little difference when the initial event is significant (such as San Diego or San Francisco events). In particular, note that the failures in the artificially attached part of Texas do not lead to cascades when the network is $(N{-}1)$-resilient. This is due to the fact that this part is connected to the whole network using small number of lines (which carry no power in normal operation in practice).  However, when the network is only $N$ resilient, these failures do propagate to the whole network\footnote{This happens also even when the FoS is $2$ (these results are not shown due to space constraints).}.

\subsection{Stochastic Outage Rule}
The third set of experiments was performed using a \emph{stochastic} outage rule as defined in (\ref{eqn:outage}) with $\varepsilon > 0$ and $p=0.5$.

\begin{figure}
\centering
\epsfig{figure=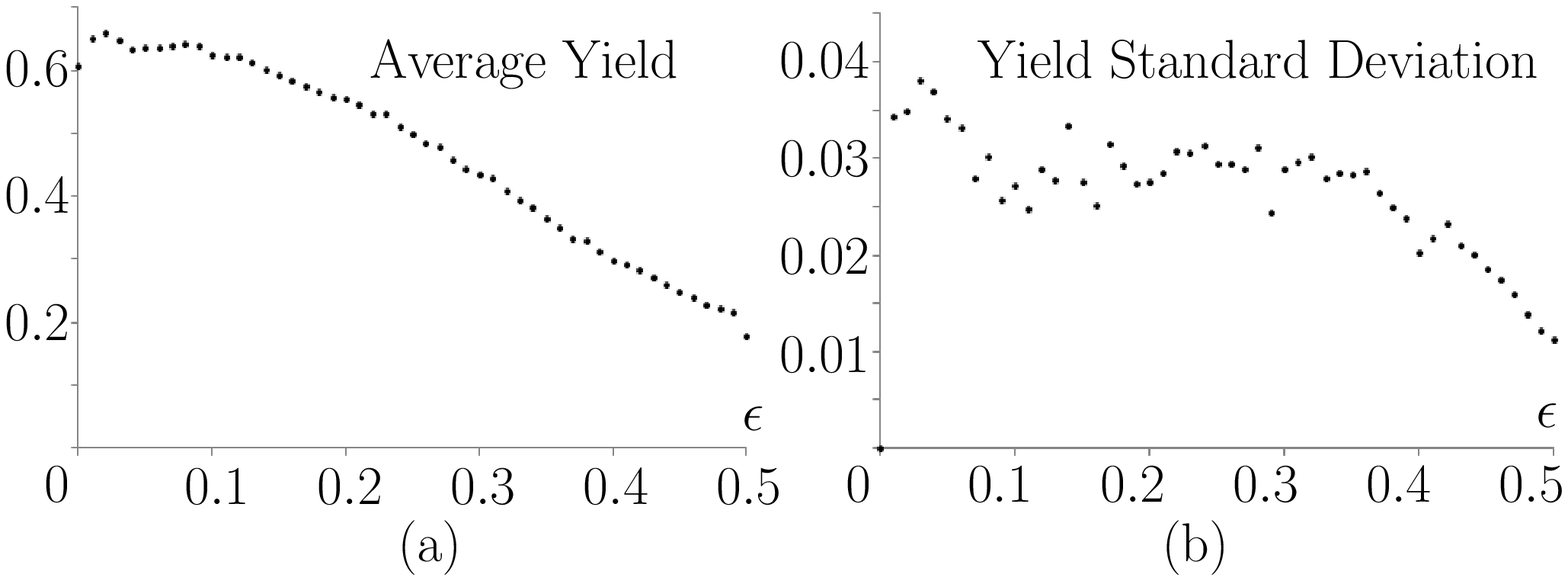, width=0.45\textwidth}
\caption{The results for a representative failure epicenter, using stochastic outage rule, based on $100$ independent runs. (a) presents the average yield, while (b) presents its standard deviation.\label{fig:yield_stoch}}
\end{figure}

In order to evaluate this outage rule, we performed two types of experiments on an $N$-resilient grid with  FoS $K=1.2$. First, for
the \emph{same} failure epicenter, we compared the yield of different values of $\varepsilon$: Figure \ref{fig:yield_stoch} shows the average yield and its standard deviation for a representative failure epicenter (the results are based on $100$ independent runs for each value of $\varepsilon$). Observe that $\varepsilon \in (0, 0.15)$ leads to a bit higher average yield than that of the deterministic rule. However, for $\varepsilon \geq 0.15$, the average yield obtained when using the stochastic rule is significantly lower.

\begin{figure}
\centering
\epsfig{figure=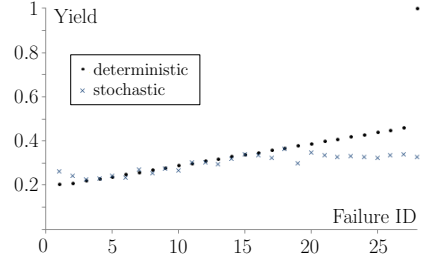, width=0.3\textwidth}
\caption{A comparison of the deterministic and stochastic outage rules for selected failure events. \label{fig:stoch_det}}
\end{figure}

In the second type of experiments,  we fixed $\varepsilon = 0.04$ and compared the results of \emph{selected} failure epicenters with the results obtained for the deterministic outage rule. The failure epicenters were chosen such that the yield using deterministic rule grows approximately linearly with the failure index. The results, depicted in Figure \ref{fig:stoch_det},  show that there is a certain yield range where the stochastic outage rule coincides with the deterministic outage rule. However, outside this range, the stochastic outage rule results in the yield values below $0.3$, which are smaller than the yield obtained by a deterministic outage rule (even when this deterministic yield is almost $1$).

%% file: sandiego.tex
\section{San Diego Blackout (Sept.~2011)} \label{sec:SanDiego}
\subsection{Description of the Blackout}
On Sept.~$8^{\mbox{th}}$, 2011, over $2.7$ million people in southwestern United States experienced a massive power blackout. Although the full details of this event are not known yet, several publicly available sources, such as \cite{SD_briefing}, make it possible to reconstruct an approximate chain of events during this blackout. As a case study, we compare the reported chain of events to our simulation results. In particular, we use this event in order to calibrate the various model parameters so that the simulation results match as closely as possible that cascade.

\begin{figure}
\centering
\epsfig{figure=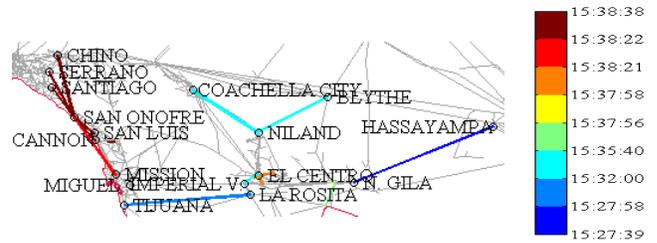, width=0.47\textwidth}
\vspace*{-1em}
\caption{The development of the San Diego blackout according to \cite{SD_briefing}.\label{fig:yuma_real}}
\end{figure}

The blackout occurred around the San-Diego county area, and involved six utility companies: San-Diego Gas and Electric Co.\ (SDG\&E), Southern California Edison (SCE), Comision Federal de Electricidad (CFE), Imperial Irrigation District (IID), Arizona Public Service (APS), and Western Area Power Association (WAPA). The power grid map of that area (using our data) is shown in Figure \ref{fig:yuma_real}\footnote{In these experiments, the \emph{actual} Western Interconnect map was used, which includes the relevant parts of Northern Baja California in Mexico.}. Specifically, there are two import generation paths into this area:
\begin{enumerate}
 \item \emph{SWPL}, which is represented by the $500$KV Hassayampa-North Gila-Imperial Valley-Miguel transmission line. This path transmits the power  generated in Palo Verdi Nuclear Generating Station  in Arizona.
 \item \emph{Path 44}, which is represented by the three $500$KV transmission lines that connect SCE and SDG\&E through the San Onofre Nuclear Generating Station (SONGS).
\end{enumerate}
In addition, there are several SDG\&E local power plants, and there is (relatively small) import of power from CFE.

Prior to the event, SWPL delivered $1370$MW, Path 44 delivered $1287$MW, and the local generation was $2229$MW~\cite{SD_briefing} (this includes the generation of both SDG\&E and CFE power plants). The cascade started at 15:27:39, when the $500$KV Hassayampa-North Gila transmission line tripped at the North Gila substation. Several sources indicate that this failure was caused by maintenance works performed at this substation at that time. Initial investigation suggested that this single line failure caused the blackout\footnote{Recently, some of the media publications mentioned the fact that this was not the only fault in that area. However, since these facts are still under investigation, we prefer to reconstruct the cascading failure according to the chain of event described in~\cite{SD_briefing}.}.  The actual cascade development is shown in Figure \ref{fig:yuma_real}. 

\subsection{Simulation Results}

We performed two sets of experiments.
 In the first set, instead of performing simulation on the entire Western Interconnect, we chose to use a part of the grid which includes only the affected area. The initial conditions were set to match as close as possible the actual conditions prior to the event. In particular, we set the generation of the Palo Verde nuclear plant (which is the main contributing import generation unit) to  $3{,}600$ MW out of its nominal $4{,}300$MW. This resulted in the following initial conditions of the  import generation: $\mbox{SWPL}= 1{,}386$MW, $\mbox{Path 44}  = 1{,}284$MW.

\begin{figure}
\centering
\epsfig{figure=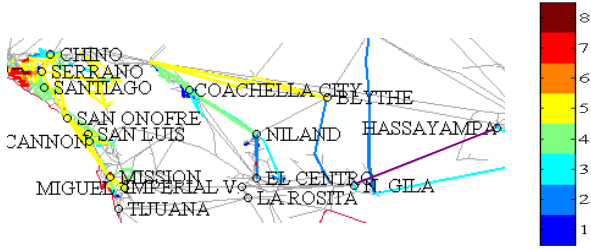, width=0.43\textwidth}
\vspace*{-1em}
\caption{The development of the San Diego blackout in the first eight rounds using our simulation.\label{fig:yuma_sim}}
\end{figure}

Moreover, since in the actual event there was no $(N-1)$-resilience with respect to the faulted North Gila--Hassyampa line, we used an $N$-resilient grid  with different values of FoS $K$ (recall Section~\ref{sec:capacities}). In addition, by \cite{SD_briefing}, the actual capacity of Path 44 is almost $2.7$ times the flow in normal operation. This information also correlates with other sources (e.g., \cite{WECC_hist}) which indicate that the power capacities are not based on a uniform FoS parameter. Since Path 44 was a major factor of the cascade development, as it carried most of the lost SWPL power, we decided to adjust its FoS accordingly. In particular, its FoS was set to $2.5$. After experimenting with the value of $K$ for other lines, we found out that $K=1.5$ leads to a behavior that most resembles that of the actual event . The resulting cascade behavior is shown in Figure \ref{fig:yuma_sim}.

Table~\ref{fig:comp} presents a brief comparison of the simulation results and the known details of the actual event. The description of the actual event is presented exactly as in \cite{SD_briefing}, without any interpretation. It can be observed that although the simulated cascade does not follow exactly the actual one, both of them developed in a similar way. This suggests that our model and data can be used to identify the vulnerable locations and design corresponding control mechanisms that will allow to stop the cascades in the early stages.

\begin{figure}
\centering
\epsfig{figure=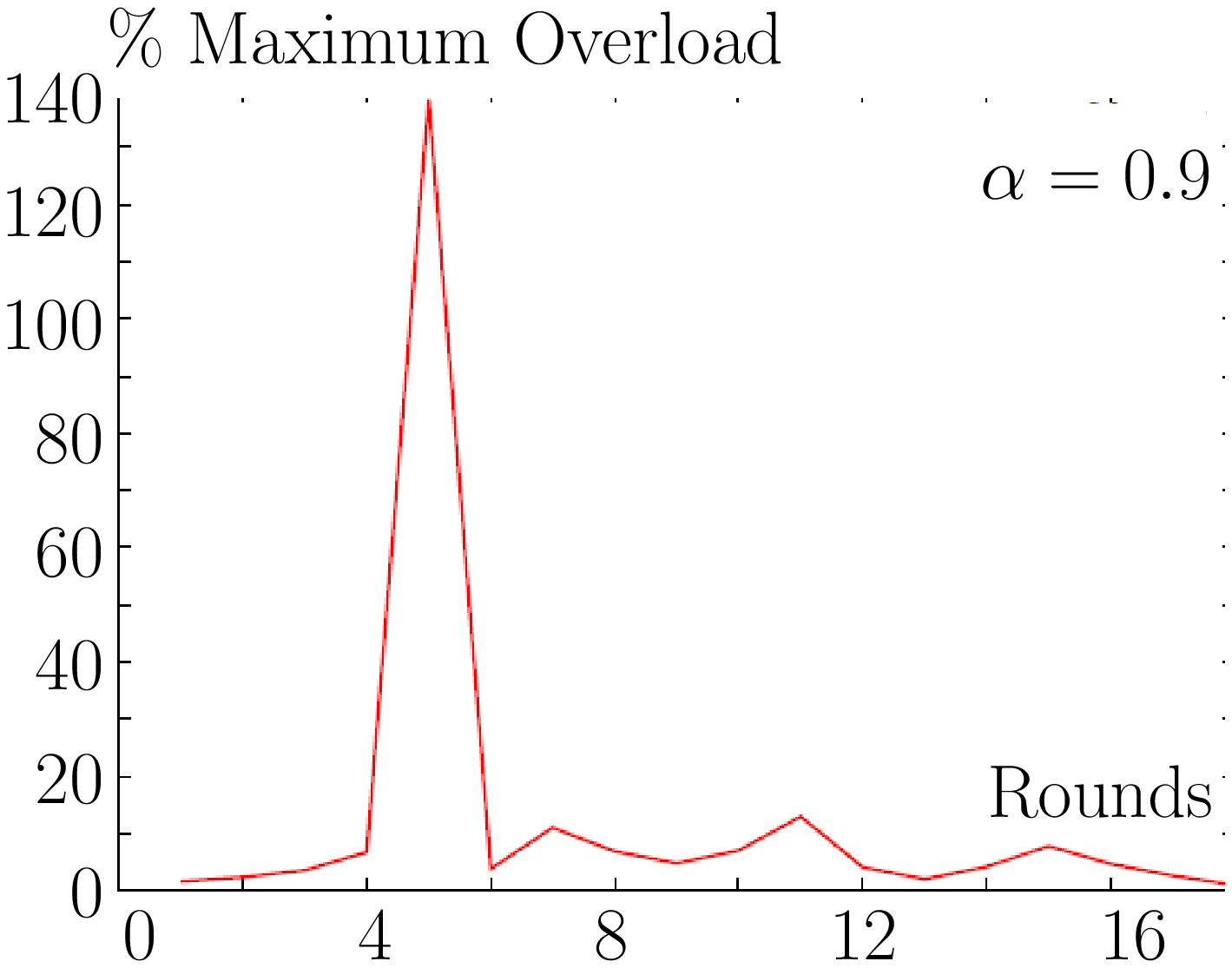, width=0.235\textwidth}
\epsfig{figure=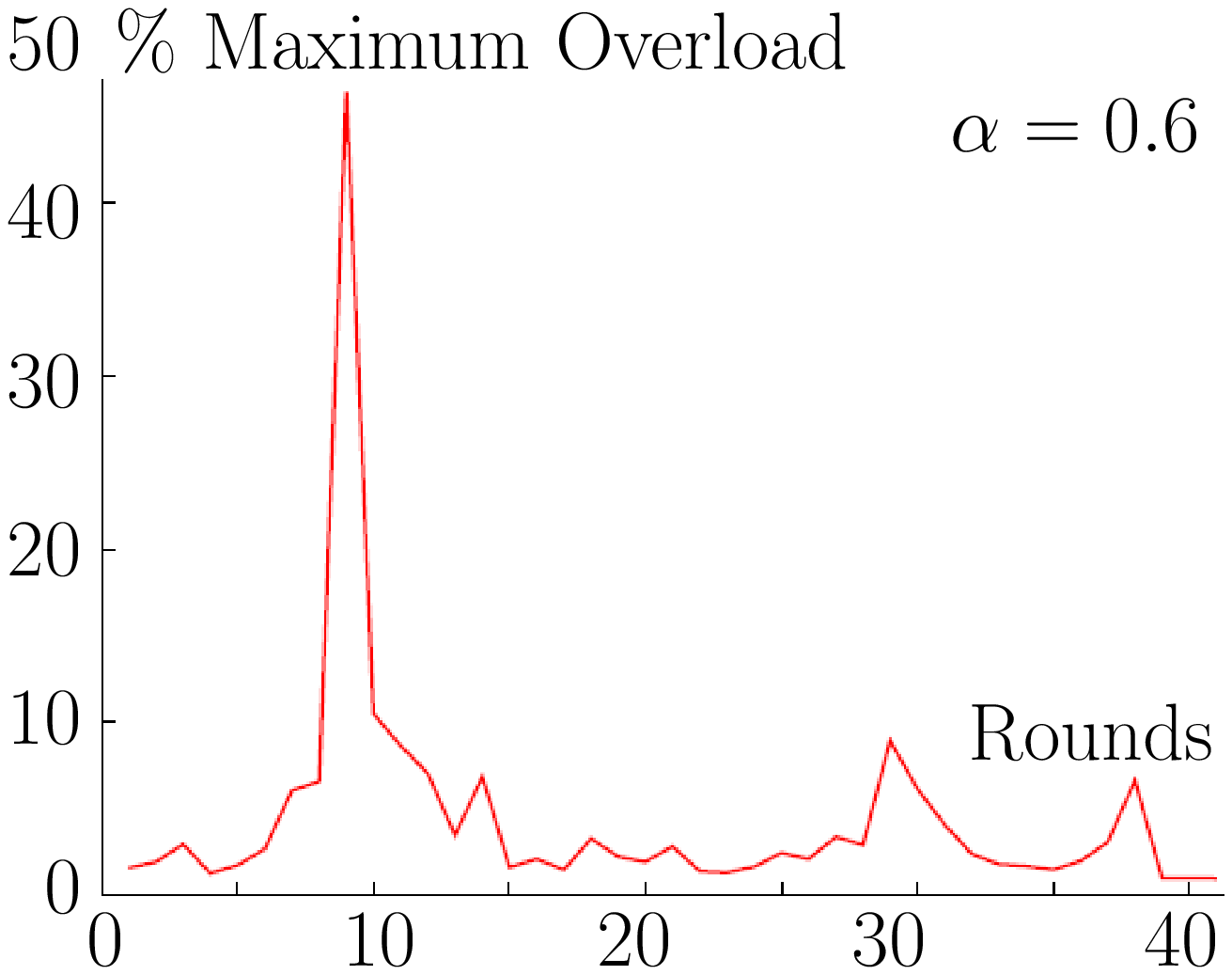, width=0.225\textwidth}\\
\epsfig{figure=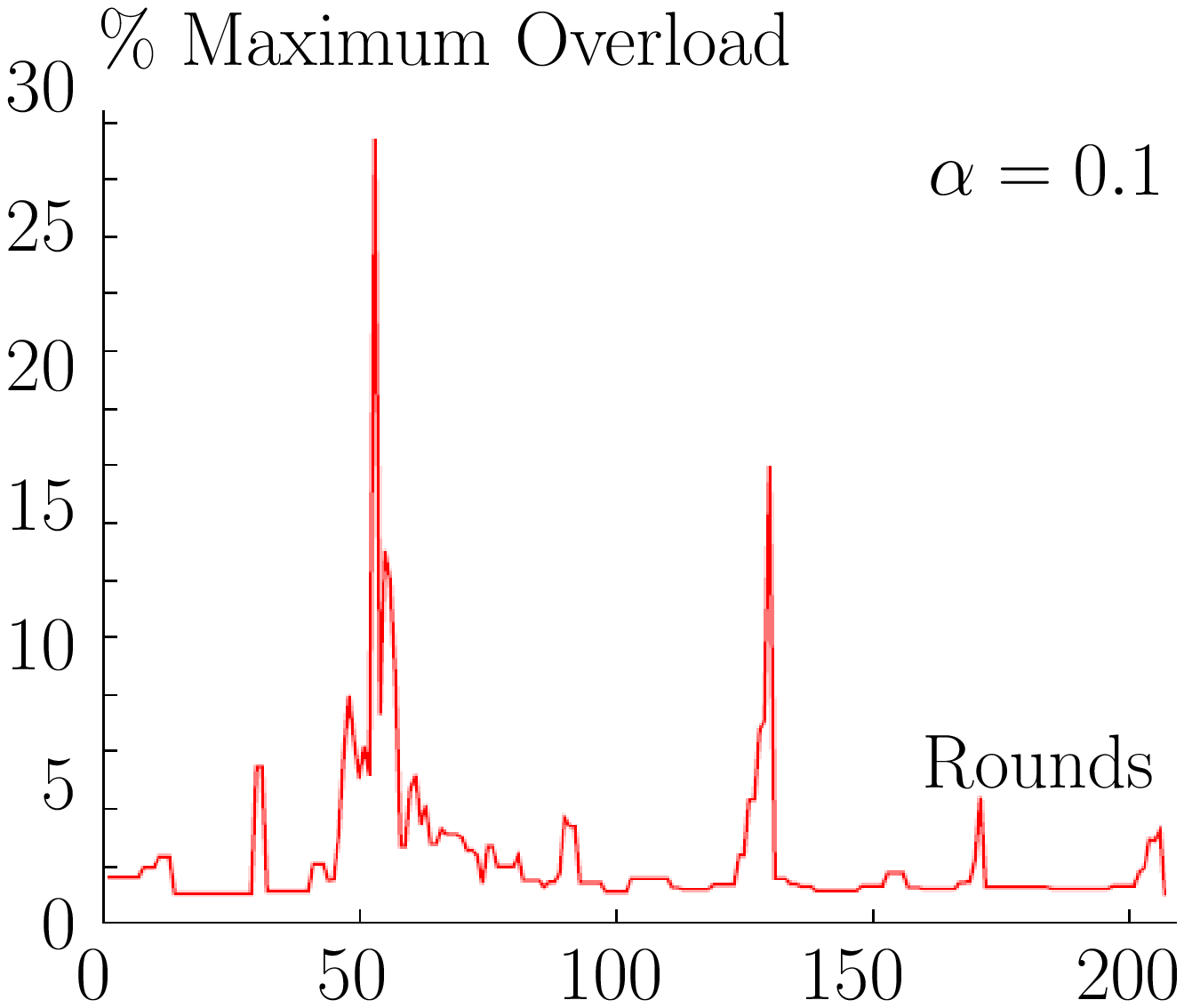, width=0.22\textwidth}
\epsfig{figure=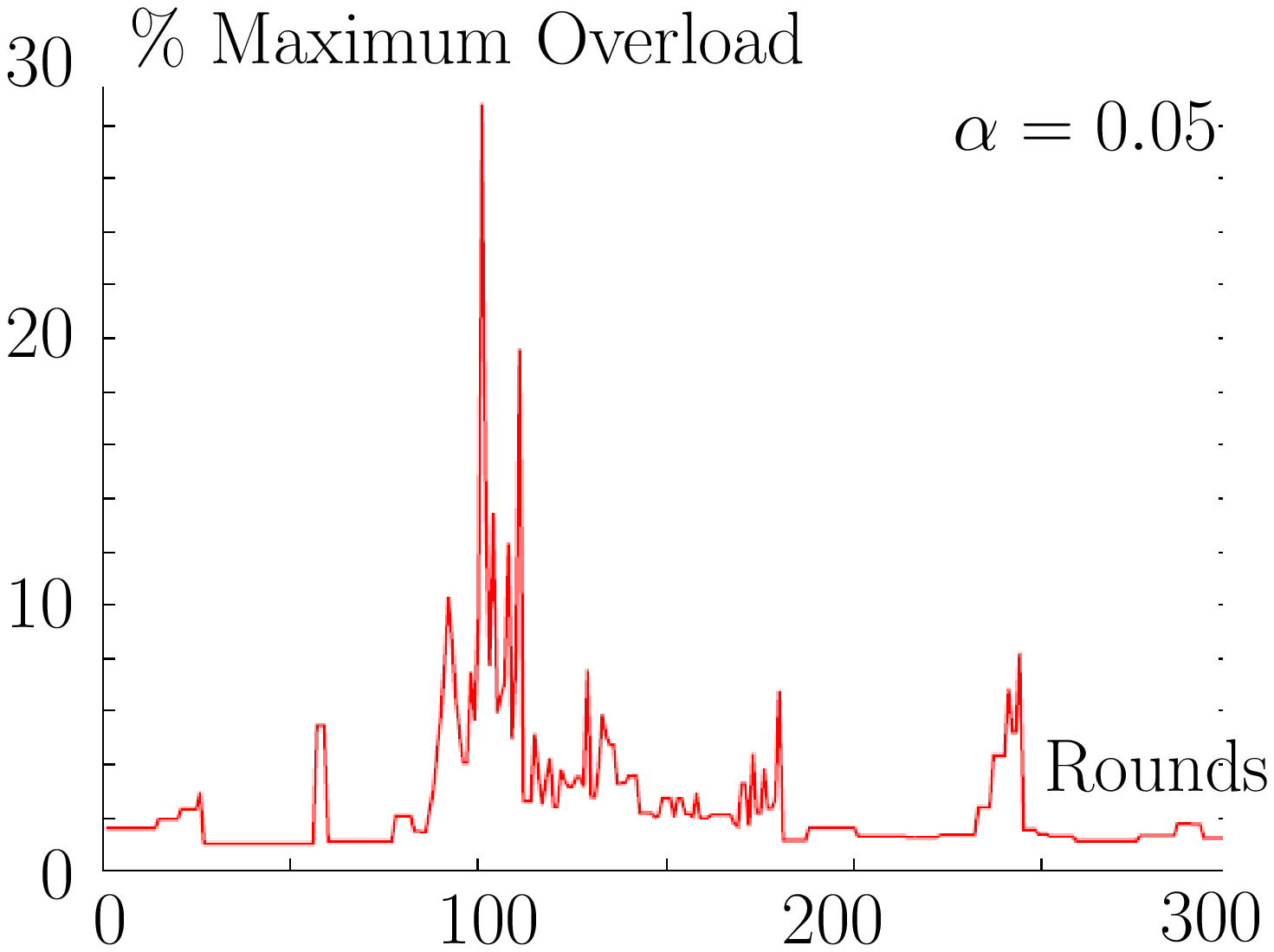, width=0.24\textwidth}
\caption{Maximum line overload for different values of $\alpha$.\label{fig:alpha}}
\end{figure}

The second set of experiments was performed on the whole Western Interconnect, and the goal was to examine the effect of the moving average parameter $\alpha$ on both the maximum line overloads and  the length of the cascade. The results (see Figure \ref{fig:alpha}) show that the larger $\alpha$ is, the higher is the maximum load and the shorter is the cascade. 
Moreover, when $\alpha$ is small (i.e., less than $0.5$), there is a period of time when the maximum overload is smaller than that at the initial round. This suggests that a control mechanism that is applied at that time will stop the cascade with relatively high yield in early rounds.

\begin{table}
\centering
{\scriptsize
\begin{tabular}{|c|l||c|l|}
 \hline
\multicolumn{2}{|c||}{Actual} & \multicolumn{2}{|c|}{Simulation} \\
\hline
\hline
& & 1 & Path 44: $1415$MW; \\
& &    & El Centro substation internal \\
& &    & line overload of ~$100$MW. \\
\hline
15:27 & Path 44: $2407$MW; & 2 & Path 44: $1438$MW; \\
& Problems with Imperial &  & El Centro internal line trip.\\
& Valley-El Centro line & &\\
&  resulting in 100MW & &\\
&  swing. & &\\
\hline
15:32 & Path 44: $2616$MW; & 3 & Path 44: $1992$MW; \\
& Two lines trip at & & \\
& Niland-WAPA and  & & \\
& Niland-Coachella Valley. & & \\
\hline
15:35 & Path 44: $2959$MW; & 4 & Path 44: $3043$MW; \\
& IID and WAPA are & &  Niland-Coachella Valley \\
& separated. & &  line overload. \\
\hline
15:37 & Path 44: $3006$MW; & 5 & Path 44: $2991$MW; \\
& IID tie line to WAPA & & Niland-WAPA and Niland- \\
& trips.  & & Coachella Valley lines trip. \\
\hline
15:38 & Path 44 trip; & 6 & Path 44 trip; \\
& SONGS trips. & & 4 out of 7 lines from SONGS \\
&   & & to San Diego trip. \\
\hline
&                        & 7 & SONGS stabilizes with total\\
& & &  generation of $1350$MW \\
& & &  out of $2253$MW. \\
\hline
 \end{tabular}
}
\caption{Comparison of the actual event and the simulation results.  \label{fig:comp}}
\end{table}

%% file: control.tex
\section{Control} \label{sec:control}
This section describes experiments with control algorithms in the specific case of the San Diego event, illustrated in Figure \ref{fig:cascades_5_1}(a). The general goal of such algorithms is to stop the cascade (and, if possible, in a short time frame) without
losing much demand.

In particular, we consider an algorithm that
will shed a minimum amount of demand so as to yield a stable grid --
the cascade has been stopped.  In this paper (for lack of space), we
focus on algorithms that will operate within a single round of the
cascade. The critical question is, then, {\em at which round} control
should be applied.  Assuming that a given round $t$ is under consideration, our control is constrained as follows:
\begin{itemize}
\item [(a)] At each demand point $i\in\mathcal{D}$,  we reduce the demand by a certain quantity, $s_i$.
\item [(b)] We adjust generator output, within each component (a.k.a.~island) so as
to maintain overall balance.
\item [(c)] However, generators are furthermore constrained in that the amount
of change in a generator must be proportional to its current output.
\item [(d)] After the demand shedding and generator adjustments,
the moving-average power flow, on each operating line $(i,j)$, cannot
exceed its capacity $u_{ij}$.
\end{itemize}
Rule (c) approximates generator ``ramp-up'' and ``ramp-down''
constraints (broadly speaking, generators cannot modify their output arbitrarily fast).  Rule (d) states that, according to our thermal model, the cascade will stop.  Rules (a)-(d) describe our constraints; the goal is to pick the round $t$ and
the quantities $s_i$ so as to maximize the remaining demand.

Note that, at a given round $t$, this optimization problem can be
written as a linear program. Specifically, denote by $\tilde
f_{ij}^t, \tilde
D_i^t, \tilde P_i^t$ the value, just before round $t$, of the flow on line $(i,j)$, the demand at demand point $i\in\Dl$, and
the generation at supply point $i\in\Cl$, respectively. Furthermore,
denote by $C_1,\ldots C_n$ the connected components in the power grid
graph before round $t$, and let $\mbox{comp}(i)$ denote the connected
component that contain node $i$.  The linear program is as follow:

\small
{\[
\begin{array}{ll}
 \mbox{\bf minimize } \sum_{i \in \Dl}s_i \mbox{ }\mbox{\bf subject to } & \\ & \\
 0\leq s_i \leq \tilde{D}_i^t & \forall i\in \Dl \\
 \alpha | f_{ij} | \ + \ (1 - \alpha) \tilde f_{ij}^t \ \le \ ( 1 -
 \varepsilon) u_{ij} & \forall \mbox{ line } (i,j) \\
\sum_{(i, j) \in \delta^+(i)} f_{ij}{-}\sum_{(j, i) \in \delta^-(i)}
f_{ji}{=}P_i, & \forall i \in \Cl \\
\sum_{(i, j) \in \delta^+(i)} f_{ij}{-}\sum_{(j, i) \in \delta^-(i)}
f_{ji}{=} {-}(\tilde{D}_i^t{-}s_i) & \forall i \in \Dl \\
\sum_{(i, j) \in \delta^+(i)} f_{ij}{-}\sum_{(j, i) \in \delta^-(i)}
f_{ji}{=}0, & \forall i\in \Nl{\setminus}(\Cl{\cup}\Dl) \\
\theta_i - \theta_j - x_{ij}f_{ij} = 0 & \forall \mbox{ line } (i,j)
\\
0\leq \lambda^{C_m} \leq 1 & \forall \mbox{ component } C_m \\
P_i = \tilde{P}_i^t (1 - \lambda^{\mbox{comp}(i)}) & \forall i\in\Cl
\\
\sum_{i \in C_m{\cap}\Cl} P_i = \sum_{i \in C_m{\cap}\Dl} D_i &\forall \mbox{ component } C_m \\
\end{array}
\]
}
\normalsize
\noindent{where,} as in Section~\ref{sec:power_mod}, $\delta^+(i)$
($\delta^-(i)$) is the set of lines oriented out of (into) node
$i$. Notice that third-sixth equations in the linear program above are
identical to Equations~(\ref{eqn:flow1})-(\ref{eqn:flow2}) in Section~\ref{sec:power_mod}.

\begin{figure}
\centering
\epsfig{figure=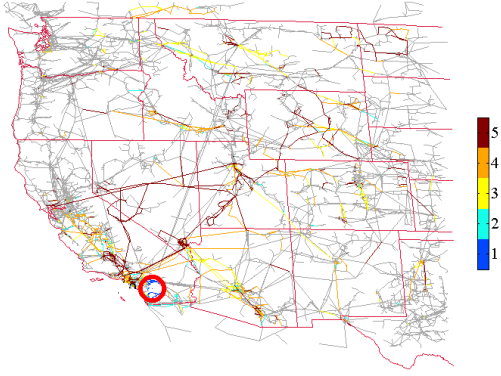, width=0.37\textwidth}
\caption{Illustration of cascading failures over $5$ rounds using stochastic outage rule with $\varepsilon = 0.05$, $p = 0.5$, and $\alpha = 0.1$. The colors represent the rounds in which the lines faulted. \label{fig:cascades_stoch}}
\end{figure}

As mentioned before, we demonstrate our control mechanism by considering a failure event in San
Diego area. Figure \ref{fig:cascades_stoch} presents
the development of this event over the first five rounds of a particular run obtained using the stochastic
line failure model with $\varepsilon = 0.05$, $p = 0.5$, and $\alpha =
0.1$ (that is, a simulation with small time increments).

Table~\ref{controltable} outlines the performance of the optimal
control mechanism, where ``Round'' refers to the round on which the
optimal control is applied,
while ``Yield'' refers to the outcome.  We
see from the table that applying control at the outset of the cascade
is {\em not} optimal (this is typical, in our experience). On the
other hand, waiting too long is not optimal either.  Rather, there is a critical frame of time where effective
control is possible; the precise time frame can be discovered
by running our simulation upon the failure event, and applying the
control only when we reach the round with optimal outcome. We also
note that without control, the cascade stops at round $74$ with the
yield of $0.34$. Currently, we are developing robust versions of this algorithm with
respect to errors in data, timing, and delays in implementation.

\begin{center}
\begin{table}[tb]
\caption{Optimal control outcome.} \label{controltable}
\centering
\begin{tabular}{|l| c c c c c c c c|}
\hline
\small Round  &    1 &    5 &  10 &    20 & 30 & 40 & 50   & 74 \\
\hline
\small Yield & 0.22 & 0.55 & 0.49 & 0.41 & 0.39& 0.38 & 0.36  & 0.34\\
\hline
\end{tabular}
\end{table}
\end{center}

%% file: conc.tex
\section{Conclusion and Future Work} \label{sec:conc}
In this paper, we considered a DC power flow and an accompanied cascading failure model. We showed analytically that these models differ from previously-studied model based on an epidemic-like failures (which are often analyzed using percolation theory). Then,
we used techniques from optimization and computational geometry along with detailed GIS data to develop a method for identifying the power grid locations that are vulnerable to geographically correlated failures. We performed extensive numerical experiments that show the relations between the various parameters and performance metrics. Specifically, we used a recent major blackout event in San Diego area as a case study to calibrate different parameters of the simulations. We also demonstrated that the use of control at the right point in the cascade can mitigate the effects of a large scale failure. While the presented  results are for an intentionally modified version of the US Western Interconnect, they demonstrate the strength of our tools and provide insights into the issues affecting the resilience of the grid. These results can be used when designing new power grids, when making decisions regarding shielding or strengthening existing grids, and when determining the locations for deploying metering equipment.



This is one of the first steps towards an understanding of the grid resilience to large scale failures. Hence, there are still many open problems.
For example, we plan to study the results' sensitivity to the failure model (e.g., to consider a rule under which the probability of a line failure is a function of the overload). Moreover, we plan to study the effectiveness of some of the current control algorithms and their capability to cope with geographically correlated failures. Finally, we plan to develop control algorithms that will mitigate the effects of such failures and  network design tools that would enable to construct resilient grids.

%
%

%% file: appendix.tex
\appendix

\paragraph*{Proof of Observation~\ref{lem:samesum}}
Let $f_{e_i}$ be the flow along line $e_i$, let $x_{e_i}$ be its reactance and let $\theta_{v_i}$ be the
phase angle of node $v_i$. By summing over the equalities of (\ref{eqn:flow2})
for the lines of path $p_1$ we get $\theta_a-\theta_b=\sum_{e_i\in p_1}
f_{e_i}x_{e_i}$. Similarly, for $p_2$, $\theta_a-\theta_b=\sum_{e_i'\in p_2}
f_{e_i'}x_{e_i'}$, and the claim follows.

\paragraph*{Proof of Lemma~\ref{thm:arbitrarilyfar}}
Consider an $M$-ring, and suppose that $M$ is even (similar arguments can be used when $M$ is odd). Assume that the capacity of all lines is $1$, except lines $(M/2, M+M/2+1)$ and
$(M/2, M+M/2+1)'$ whose capacity is $0.5$.  Assume also that
$\alpha>0$. Initially, the ring operates flawlessly, and all power
flows are within the capacity of the corresponding lines.

Suppose that an initial failure event occur in lines $(0,M)$ and
$(0,M)'$ (that is, a parallel lines failure). As described above, all lines will either carry flow of $0$
or of $1$, and therefore all lines except $(M/2, M+M/2+1)$ and
$(M/2, M+M/2+1)'$ will continue operating normally.

As for lines $(M/2, M+M/2+1)$ and $(M/2, M+M/2+1)'$, their
post-failure power flow is $1$ while their pre-failure flow is $0.5$. Since
for each $\alpha>0$, $\alpha\cdot 1 + (1-\alpha)\cdot 0.5>0.5=u_{M/2,
  M+M/2+1}$, these lines fault in the next round. Thus, the distance
between consecutive failures in this cascade is $\Theta(M)$. As one can
choose an arbitrarily large $M$-ring, the distance between there two
consecutive failures can be made arbitrarily large and the claim
follows.

\paragraph*{Proof of Lemma~\ref{thm:arbitrarilylarge}}
When the power capacity $u$ of each line is $0.5$, the same failure
described in the proof of Lemma~\ref{thm:arbitrarilyfar}, which starts
with 2 lines failure, causes an outage of 3/5 of the lines. Notice
that in this case after the first iteration there is a demand
shedding of half the total demand. Only odd lines still operate and
each one of them carries a power flow of $0.5$ unit (which is below its
power capacity). Thus, this specific failure event stops after one
iteration.

\paragraph*{Proof of Lemma~\ref{thm:nonmonotone1}}
Consider an $M$-ring and assume that the capacity of all lines is
$0.5$, and, for ease of presentation, that $\alpha=1$. Initially, the
 ring operates flawlessly, and all power flows are within the capacity
of the corresponding lines.

Let $A=\{(0,M),(0,M)'\}$ (parallel lines failure). As described above, after a single
iteration, all tie lines and odd lines will carry power flow of $1$ unit, exceeding
their capacity and therefore faulting. This will cause a demand
shedding of $2$ unit within Area $1$ (no remaining active power lines); in
the rest of the areas there will be a demand shedding of $1$ unit (odd
demand node will be disconnected), resulting in a total demand
shedding of $M+1$, and a total yield of $(M-1)/2M<0.5$.

On the other hand, assume an area failure of Area $0$. Clearly, the
area failure is a superset of  $A$. However, the area failure event
causes only demand shedding of $2$ and therefore a yield of
$(M-1)/M>0.5$ (for every $M>2$).  The reason is
that in that case the cascade does not propagate outside Area $0$, and
all other power flows remain the same.
\paragraph*{Proof of Lemma~\ref{thm:nonmonotone2}}
Consider an $M$-ring and an $M$-ring with no tie-lines (which is a
subgraph of the $M$-ring). Assume that the capacity of all lines is
$0.5$, and, for ease of presentation, that $\alpha=1$.

As we have shown before that a failure in $\{(0,M),(0,M)'\}$ causes a yield
less than $0.5$ in the $M$-ring. On the other hand, when there are no
tie-line, this failure is contained within Area $1$, resulting in a
yield of at least $(2M-2)/2M>0.5$ (for $M>2$).  The same localization
property holds for all other types of failures.

\paragraph*{Proof of Lemma~\ref{lemma:singlefailure}}
Assume that the flow on line $(0,M)'$ is $y$. Thus,
(\ref{eqn:flow1}) and Observation~\ref{lem:samesum} imply that the flow of
each of the lines $(0,M+1)$ and $(0,M+1)'$ is
$\frac{1}{2}(2-y)=1-\frac{y}{2}$. In addition, (\ref{eqn:flow1})
implies that the flow on the tie line connecting Area 0 and Area 1 is
$1-y$.

Focus on Area 1. Similar arguments yield that the flow on $(1,M+2)$
and $(1,M+2)'$ is $y/2$, the flow on  $(1,M+2)$
and $(1,M+2)'$ is $1-y/2$ and the flow on the tie line is $1-y$. By
symmetry, the same flow assignment holds for all other areas.

Now consider two paths between node $0$ and node $M$. One traverses
the single line $(0,M)'$ that carries $y$ unit of power flow. The
other traverses $M$ tie lines, $M$ odd lines, and $M-1$ even
lines; the power flow along even lines is in the
opposite direction. Thus, the total flow along this path is
$M(1-y)+M(1-y/2)-(M-1)(y/2)= 2M - 2My + y/2$. Thus, by Observation~\ref{lem:samesum},
$y= 2M- 2My + y/2$ implying that $y = \frac{2M}{(0.5 + 2M) }$.

\paragraph*{Proof of Corollary~\ref{cor:ninth}}
The proof follows by case analysis. First, consider a failure event of
a tie line. Since there is no power flow along the tie line, such a
failure does not change the operations of the power grid and a
capacity of 0.5 along internal lines and 0 along tie lines suffices to
withstand such a failure.

On the other hand, Lemma~\ref{lemma:singlefailure} analyzes a single
internal line failures. In that case, as $M$ goes to infinity, the post-failure flow
on  an internal lines approaches $1$. On the other hand, the maximum
post-failure flow on the tie lines is $1/9$ when $M=2$.

\paragraph*{Proof of Lemma~\ref{lem:long}}
\begin{figure}
\centering
\epsfig{figure=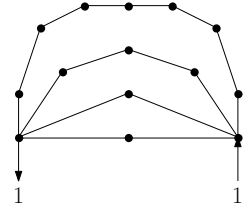, width=0.17\textwidth}
\caption{Illustration of the proof of Lemma~\ref{lem:long}. The graph
  $\mathcal{Q}_4$ has 4 paths between its generation node and its
  demand node, two of length $2$, one of length $4$, and one of length
  $8$. \label{fig:parallel}}
\end{figure}
For any $m>2$, let $\mathcal{Q}_m=\langle \Cl\cup\Dl,
\mathcal{E} \rangle$ be an undirected graph with a single supply
node and a single demand node, that are connected through $m$ disjoint
path in the following manner: the first two paths are of length $2$
(implying that along each of these path there is an additional
intermediate neutral node, connected to the generator on one side and to the
demand node on the other side). For any $i>2$, path $i$ is of length
$2^{i+1}$. See Figure~\ref{fig:parallel} depicting $\mathcal{Q}_4$. Assume also that all
lines have the same capacity $u=1/2$ and the same reactance $x=1$. We
note that the total number of lines in $\mathcal{Q}_m$ is $2^m$ and the
total number of nodes is $2^m-m+2$.

By Observation~\ref{lem:samesum}, the sum of flows along each of the $m$
path is the same, denoted by $y$. In addition, since there is no generation or demand
along each path and since each intermediate node has degree $2$,
(\ref{eqn:flow1}) immediately implies that all lines along the
same path carry the same amount of flow. Namely, the flow along the
lines of the first two path is $y/2$ and for each line of path $i$
($i>2$) the flow is $y/2^{i-1}$. Solving (\ref{eqn:flow1})
for the demand node, implies that $y/2+\sum_{i=1}^{m-1}y/2^i=1$ and
therefore $y=\frac{1}{1.5-\frac{1}{2^{m-1}}}<1$ (for any $m>2$). Since
the largest amount of flow on any line in the graph is $y/2<1/2=u$,
there power initially flow flawlessly.

Suppose now that there is a failure event in one of the lines along
the first path. Applying (\ref{eqn:flow1}) on the intermediate node
implies that there is no flow along the other line of that path. We
next show that this failure event causes a cascade that last $m$
iterations, implying that cascades can be made arbitrarily large (by
choosing larger graph $\mathcal{Q}_m$).

After each iteration $\ell$, denote by $y_\ell$ the total amount of
flow on each of the remaining path. We next show by induction that
the flow $y_{\ell+1}/2^{\ell}$ along each of the edges of the
$(\ell+1)$-th path exceeds $1/2$, implying that these lines fault in
the next iteration. In addition, for each $j>\ell+1$ the flow along all
lines of the $j$-th path is less than $1/2$, implying that
they still operate at the end of the iteration.

In the base case, (\ref{eqn:flow1})
for the demand node implies that $\sum_{i=1}^{m-1}y_1/2^i=1$, and
therefore $y_1=\frac{1}{1-\frac{1}{2^{m-1}}}>1$, thus the lines along
 the second path have power flow of $y_1/2>1/2$ and they fail. On the
 other hand, for any $m>2$, $y_1<2$, thus all lines of path $j>2$ has
 capacity of at most $y_1/4<1/2$.

Suppose now that in iteration $\ell$ all lines of the first $\ell$
paths fault. We show that iteration $\ell+1$ the lines of path
$\ell+1$, and the lines of the rest of the paths survive. The proof
follows by solving (\ref{eqn:flow1})
for the demand node with the surviving lines; namely,
$\sum_{i=\ell}^{m-1}\frac{y_{\ell+1}}{2^i}=1$, implying that
$y_{\ell+1}=\frac{1}{\frac{1}{2^{\ell-1}}-\frac{1}{2^{m-1}}}=\frac{2^{m-1}}{2^{m-\ell}-1}$. Thus,
  the flow on the lines of the $(\ell+1)$-th path is
  $\frac{y_{\ell+1}}{2^{\ell}} =
  \frac{1}{2}\frac{2^{m-\ell}}{2^{m-\ell}-1}>\frac{1}{2}$, while the
  flow on the lines of  any path of index larger than $\ell+1$ is  at most
  $\frac{y_{\ell+1}}{2^{\ell+1}}=\frac{1}{4}\frac{2^{m-\ell}}{2^{m-\ell}-1} <
  \frac{1}{2}$ (for any $\ell<m$).

Hence, the above cascade over $\mathcal{Q}_m$ lasts for $m$ iterations,
such that at each iteration the lines of one path fail. At the end of
the cascade, the demand and generation nodes are disconnected and the
yield is $0$.
\begin{figure}[t]
\centering
\epsfig{figure=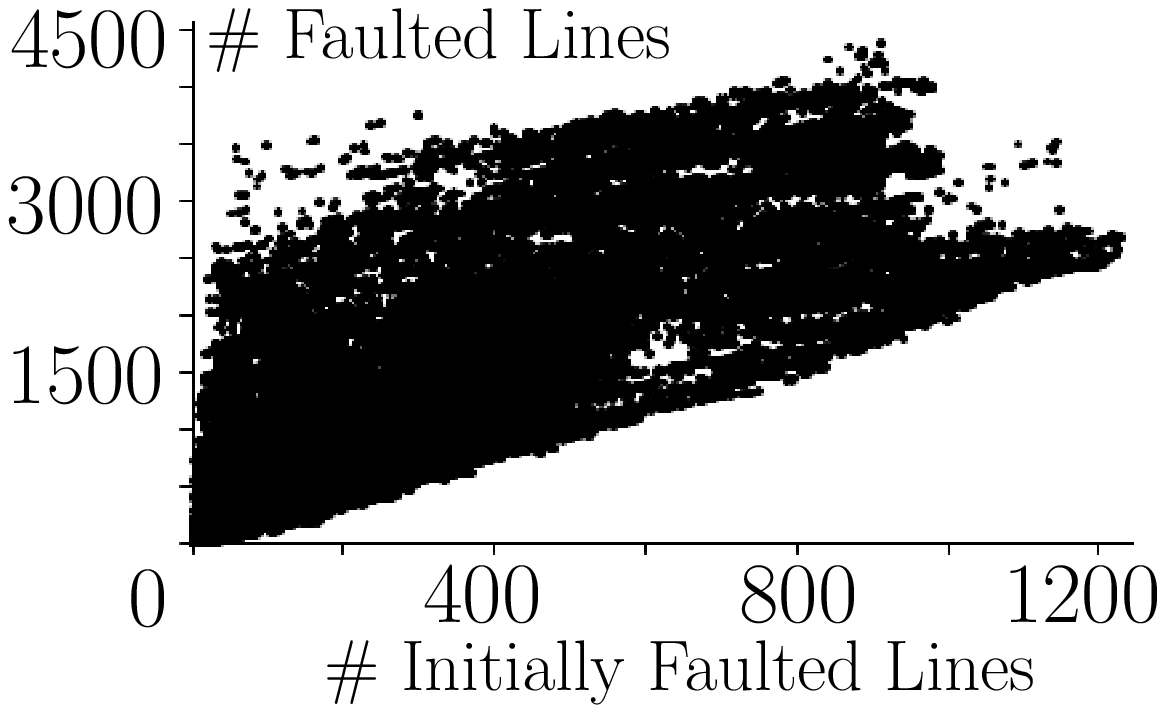, width=0.23\textwidth}
\epsfig{figure=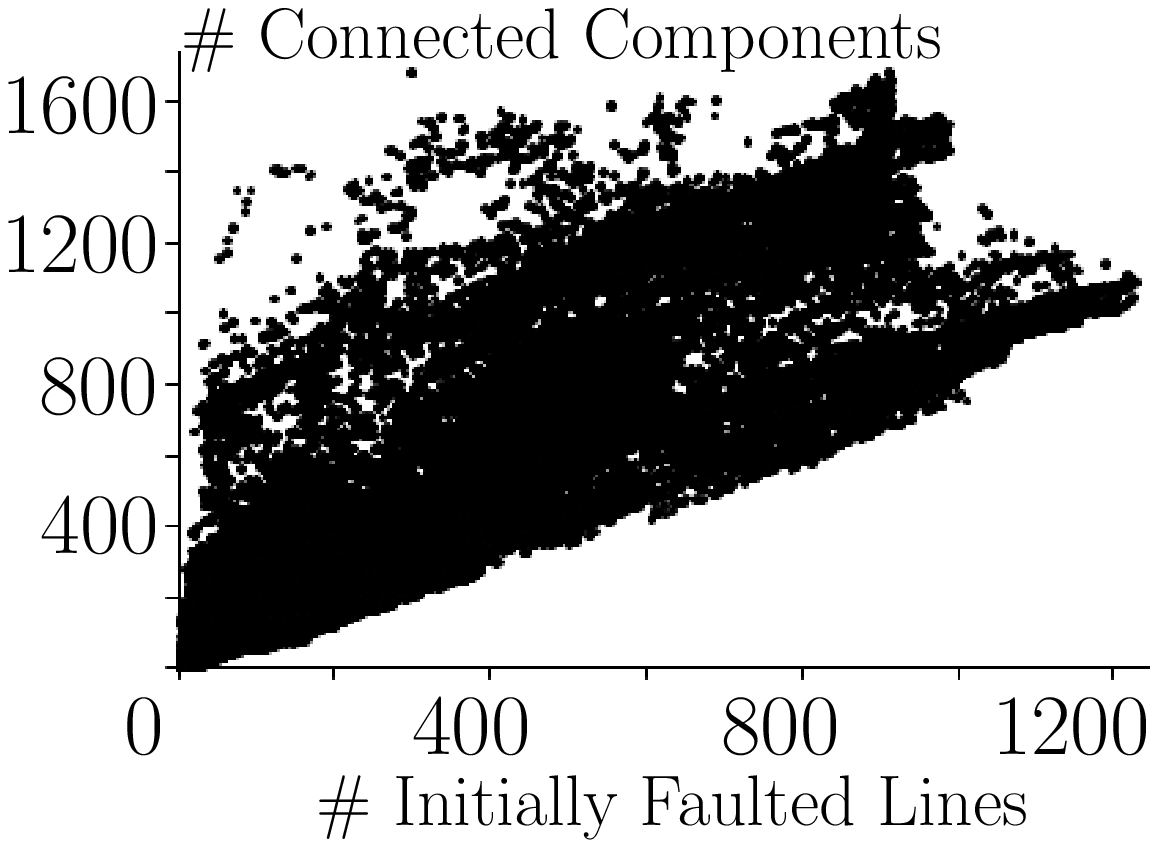, width=0.23\textwidth}
\caption{The effects of the number of initially faulted lines on the
  number of faulted lines (left) and the number of components
  (right),  after $5$ rounds of cascade for
 FoS $K=2$. \label{fig:params_5_2} }
\end{figure}

\begin{figure}
\centering
\footnotesize
\vspace*{-1em}
\subfigure[San Diego area failure event]{\epsfig{figure=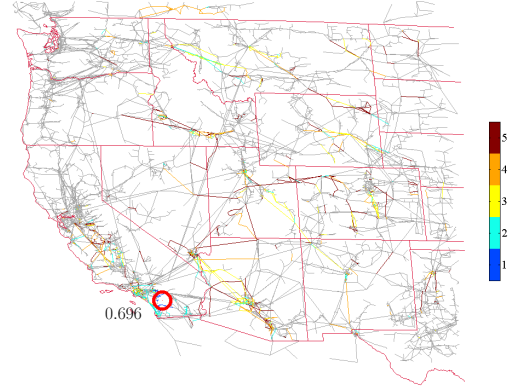, width=0.37\textwidth}} \\
\subfigure[San Francisco area failure event]{\epsfig{figure=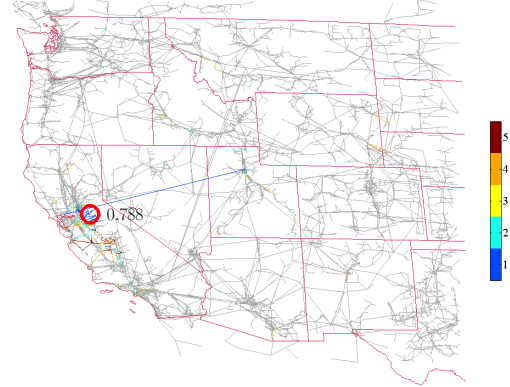, width=0.37\textwidth}} \\
\subfigure[Idaho-Montana-Wyoming border failure event]{\epsfig{figure=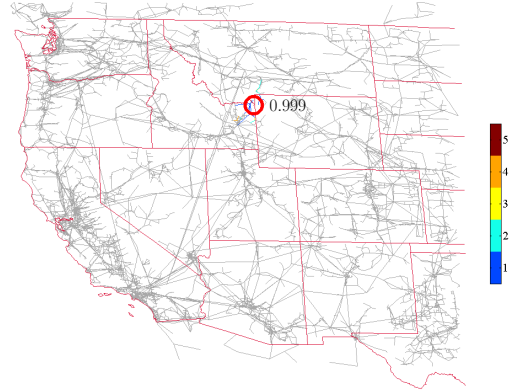, width=0.37\textwidth}}
\caption{Illustration of the cascading failures from Figure
  \ref{fig:cascades_5_1} for $N$-resilient grid with FoS $K=2$. The final yields are 0.696, 0.788, and 0.999. \label{fig:cascades_5_2}}
\end{figure}

\paragraph*{Cascades in an $N$-resilient grid with FoS $K=2$}
Figure~\ref{fig:cascades_5_2} shows the first five rounds of three
failures, initiated in three different locations. By comparing these
results to the cascade resulting from the same failures, but when the
grid has FoS $K=1.2$ (Figure~\ref{fig:cascades_5_1}), one can clearly see than increasing the FoS
significantly slows down the cascade. These results are further
verified in Figure~\ref{fig:params_5_2} which considers many failures
on that grid and shows the effects of the
number of initially faulted lines on the number of faulted lines
after 5 rounds, as well as  the number of components.

